\documentclass[reprint, amsmath, amssymb,aps,prd,reprint,superscriptaddress]{revtex4-2}
\usepackage{graphicx, color}  
\usepackage{dcolumn}   
\usepackage{bm}        
\usepackage{amssymb}   
\usepackage{amsmath}   
\usepackage{footmisc}
\usepackage{enumerate}
\usepackage{comment}
\usepackage{multirow} 
\usepackage{tabularx}
\usepackage{booktabs}
\usepackage{placeins}

\usepackage[svgnames,dvipsnames]{xcolor}
\usepackage{graphicx}
\definecolor{NewBlue}{rgb}{0.1, 0.1, 0.7}
\definecolor{NewRed}{rgb}{0.7, 0.1, 0.1}
\usepackage[colorlinks,
linkcolor=Maroon,
citecolor=NewBlue,
urlcolor=ForestGreen]{hyperref}
\usepackage{amsthm}
\usepackage[capitalise]{cleveref}
\usepackage{empheq}
\usepackage[notrig]{physics}

\usepackage[normalem]{ulem}

\usepackage{hyperref}
\usepackage{siunitx}
\usepackage{multirow} 

\newcommand{\RN}[1]{%
  \textup{\uppercase\expandafter{\romannumeral#1}}%
}

\newcommand{\ee}{\mathrm{e}}
\newcommand{\ii}{\mathrm{i}}

\newcommand{\eqdef}{\coloneqq}

\renewcommand{\Im}{\operatorname{Im}}

\newcommand{\Propag}{\boldsymbol{\mathcal{P}}}
\newcommand{\Rot}{\boldsymbol{R}}
\newcommand{\RP}{\boldsymbol{\mathcal{K}}}
\newcommand{\Xout}{\boldsymbol{x}^\text{out}}
\newcommand{\Xin}{\boldsymbol{x}^\text{in}}
\newcommand{\Wmat}{\boldsymbol{W}}
\newcommand{\vect}[1]{\boldsymbol{#1}}
\newcommand{\SR}{_\text{SR}}

\begin{document}

\preprint{APS/123-QED}
\title{Performance of multiple filter-cavity schemes for frequency-dependent squeezing in gravitational-wave detectors}

\author{Jacques Ding}
\email{ding@apc.in2p3.fr}
\affiliation{Universit\'e Paris Cit\'e, CNRS, Astroparticule et Cosmologie, F-75013 Paris, France}
\affiliation{Corps des Mines, Mines Paris, Universit\'e PSL, France}

\author{Eleonora Capocasa}
\affiliation{Universit\'e Paris Cit\'e, CNRS, Astroparticule et Cosmologie, F-75013 Paris, France}
\author{Isander Ahrend}
\affiliation{Universit\'e Paris Cit\'e, CNRS, Astroparticule et Cosmologie, F-75013 Paris, France}
\author{Fangfei Liu}
\affiliation{Universit\'e Paris Cit\'e, CNRS, Astroparticule et Cosmologie, F-75013 Paris, France}
\affiliation{School of Physics and Astronomy, Beijing Normal University, Beijing, China}
\author{Yuhang Zhao}
\affiliation{Universit\'e Paris Cit\'e, CNRS, Astroparticule et Cosmologie, F-75013 Paris, France}
\author{Matteo Barsuglia}
\affiliation{Universit\'e Paris Cit\'e, CNRS, Astroparticule et Cosmologie, F-75013 Paris, France}

\date{\today}

\begin{abstract}
Gravitational-wave detectors use state-of-the-art quantum technologies to reduce the noise induced by vacuum fluctuations, via injection of squeezed states of light. Future detectors, such as Einstein Telescope, may require the use of two filter cavities or a 3-mirror coupled filter cavity to achieve a complex rotation of the squeezing ellipse, in order to reduce the quantum noise over the whole detector bandwidth. In this work, we compare the theoretical feasibility and performances of these two optical layouts and their resilience with respect to different degradation sources (optical losses, mismatching, locking precision), analytically and numerically. We extend previous analysis on squeezing degradation and find that the coupled cavity scheme provides similar or better performances than the two-cavity option, in terms of resilience with respect to imperfections and optical losses. We further highlight the role of mode-mismatch phases in limiting squeezing. Finally, we propose a possible two-step implementation scheme for Einstein Telescope using a single filter cavity that can be possibly upgraded into a coupled filter cavity. 
\end{abstract}

\maketitle

\section{Introduction}
Over the past decade, LIGO, Virgo and KAGRA have opened the era of gravitational-wave (GW) astronomy, detecting hundreds of signals from merging compact binary objects, including black-hole binaries, neutron-star binaries and black-hole-neutron-star systems \cite{PhysRevX.13.041039,gwtc4}. These observations have led to a wide range of scientific insights in fields as diverse as general relativity, astrophysics and cosmology \cite{LVKresults}.

\begin{figure}
    \centering
\includegraphics[width=0.7\linewidth]{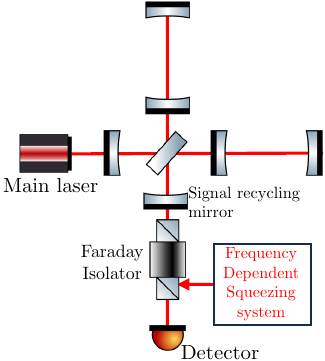}
    \caption{Simplified optical scheme for dual-recycled GW interferometer with Frequency-Dependent Squeezing injected through the output port via Faraday cirulator/isolator. Current and future generations of GW detectors use the same simplified scheme, but ET-LF operates in a detuned configuration. Here, for clarity, we represent the arms at a right angle.}
    \label{fig:ITF}
\end{figure}

Quantum noise limits the ultimate sensitivity of GW detectors, due to vacuum fluctuations of the electromagnetic field entering the detector through the output port. It can be mitigated by injecting squeezed vacuum through this port \cite{Caves_SQZ,Barsotti_2019}. Because of the inherent frequency response of the interferometer, and the ponderomotive coupling with the mirrors, such a squeezed vacuum state needs to acquire a frequency dependence in order to optimally reduce quantum noise in the whole detector bandwidth. Precisely, the interferometer rotates the squeezed quadratures in a frequency-dependent manner \cite{klmtv}.  \cref{fig:ITF} shows the simplified optical scheme of such a quantum-enhanced GW interferometer.

Frequency-Dependent Squeezing (FDS) is achieved by first generating a frequency-independent squeezed vacuum state, then passing it through one or several filter cavities to rotate its quadratures, thus compensating the response of the interferometer \cite{virgo_FDS,zhao_2020,Oelker_2016,McCuller_2020}. Recently, the Advanced LIGO detectors achieved FDS using a single filter cavity \cite{aligo_FDS}, which improved the detection rate up to 65\% \cite{Capote_2025} and allowed continuous operation below the standard quantum limit \cite{Jia_2024}.

At present, “third-generation” (3G) detectors, to be built in new infrastructures, are being studied and aim to increase sensitivity by a factor of $\sim 10$ compared to existing detectors. They will continue the scientific program of LIGO-Virgo-KAGRA, addressing fundamental open questions such as the nature of gravity and dark energy, the properties of nuclear matter, and the formation of neutron stars and black holes throughout cosmic history \cite{abac2025scienceeinsteintelescope}. Einstein Telescope, one of such 3G detectors, is composed of two interferometer designs: a “room temperature” detector, optimized for high-frequency sensitivity, and a “cold” detector (ET-LF), operating at cryogenic temperatures for improved low-frequency sensitivity \cite{ET_design_update}. This work focuses on the quantum noise reduction (QNR) in ET-LF for each 10-km, detuned, dual-recycled Fabry-Pérot Michelson interferometer, keeping the results as general as possible and independent on the exact angle between the arms. We shall only particularize to the triangular geometry currently proposed \cite{ET_design_update} for Einstein Telescope when computing the astrophysical reach.

ET-LF is expected to operate in a detuned (non-broadband) configuration \cite{Buonanno_2001,purdue2002practical}. This so-called ``detuning" corresponds to a constant, $\mu m$-scale offset of the position of the signal recycling mirror (see \cref{fig:ITF}) away from the resonant length for the carrier field. Physically, this induces a resonant extraction of the GW signal around a frequency determined by the detuning, thus tailoring the detector to be more sensitive to frequency ranges of interest. Due to this complex response of the interferometer, the corresponding optimal FDS scheme requires strictly more than a single filter cavity; usually two (referenced as 2FC in the following \cite{Harms_2003}) are needed.

Among alternatives schemes that do not require two physical filter cavities, such as EPR squeezing \cite{Ma_2017,yap_2020,Sudbeck_2020} and quantum teleportation \cite{Nishino_2024}, we consider here a three-mirror cavity, otherwise known as a coupled filter cavity (henceforth referenced as CFC), which has been first proposed in \cite{Jones_2020}. A comparison of 2FC vs. CFC in the lossless case was performed in  \cite{Zhang24}, where it was identified that the equivalence between the two schemes imposed a value for the CFC middle mirror transmissivity apparently smaller than current state of the art techniques in mirror coating. 

In this paper, we review the equivalency between 2FC and CFC, and  address the previous issue of the middle mirror transmission (\cref{sec:configs}). In particular, we show that, for longer filter cavities than that studied in \cite{Zhang24}, its value is actually readily achievable; we find that transmission errors can be compensated by adjusting the cavities' detunings. We then perform a complete comparison between 2FC and CFC configuration in terms of squeezing degradation budget, in particular including the effects of optical loss, mismatch and phase noise (\cref{sec:comp}). It is shown that, theoretically, CFC may provide better resilience to mode-matching and phase noise than 2FC. Finally, we introduce a tuned configuration of ET-LF (with single filter cavity) as an intermediary step before achieving detuned ET-LF, and study its astrophysical merit (\cref{sec:applications}).

\section{Proposed configurations}
\label{sec:configs}

\begin{figure}
    \centering
    \includegraphics[width=\linewidth]{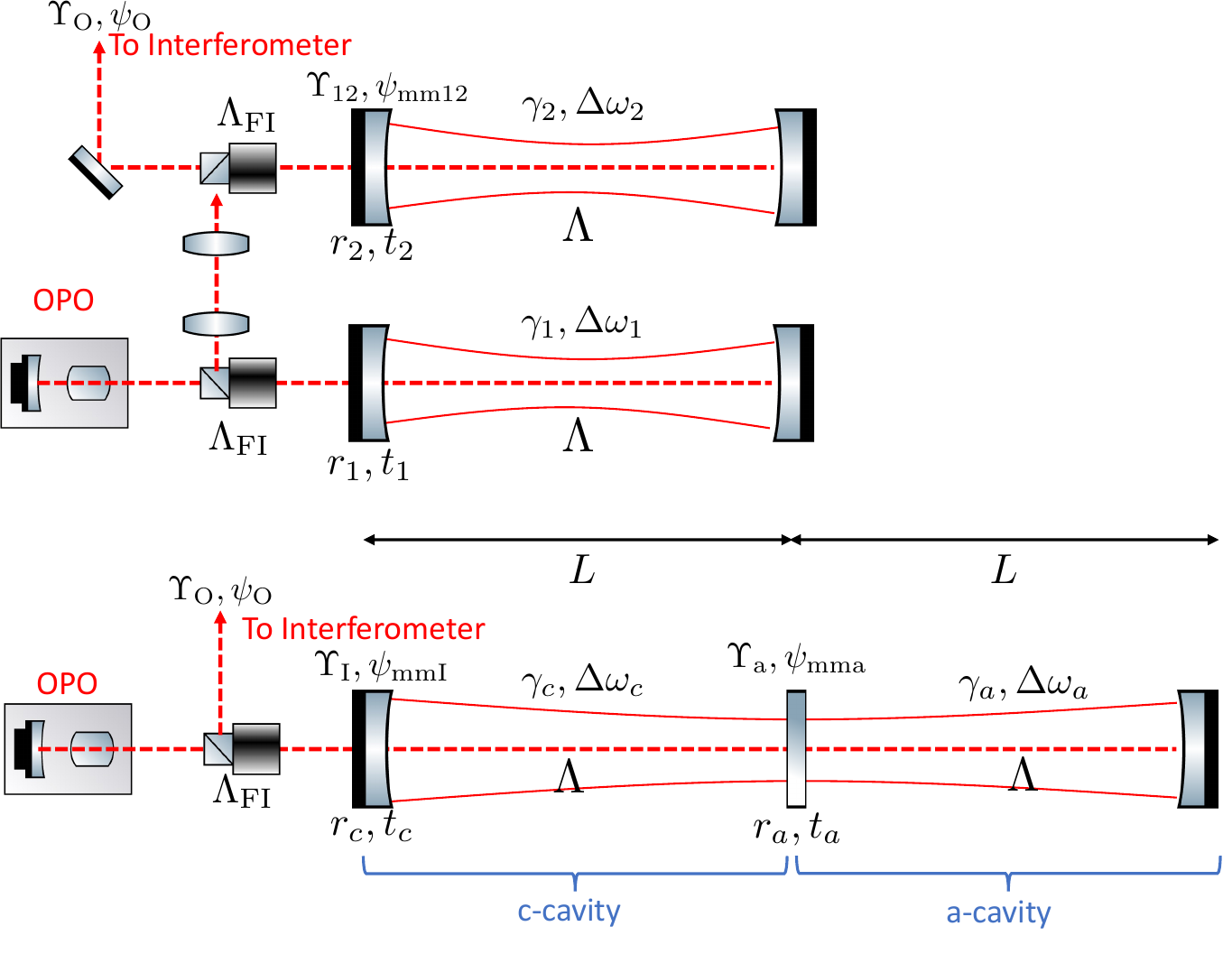}
    \caption{Optical schemes of the two-filter cavity (2FC, top) and coupled filter cavity (CFC, bottom) configurations. Note that, aside from the different cavity configuration, the 2FC differs from the CFC insofar at is requires mode matching lenses between the first filter cavity (FC1) and the second one (FC2), as well as an additional Faraday Isolator (FI).}
    \label{fig:2FC_CFC_optical_scheme}
\end{figure}

We review below the two optical configurations for producing FDS which are the subject of this paper. Let $L_\text{tot}$ be the total length of the filter cavities. They are shown in \cref{fig:2FC_CFC_optical_scheme}.
\begin{itemize}
    \item 2FC configuration with cavity lengths $L_1$, $L_2$ such that $L_1 + L_2 = L_\text{tot}$.
    \item CFC configuration with cavity lengths $L_c$, $L_c$ such that $L_a + L_c = L_\text{tot}$; the first sub-cavity is called "c-cavity" and the second is the "a-cavity". 
\end{itemize}

Note that the main difference between these two configurations, apart from the cavities, is the additional Faraday Isolator (FI) as well as free-space mode matching elements between the first and second cavities in 2FC. These elements will further degrade squeezing compared to CFC as we will show in \cref{sec:comp}. 

\medskip

\subsection{Scaling law between 2FC and CFC}
In the lossless case and assuming that all the cavities considered have high finesses, it is possible to obtain closed form equations relating the linewidths and the detunings of 2FC to those of CFC. They are given in \cite{Zhang24} and recalled here:
\begin{align}
\label{eq:gamma_c}
    \gamma_c &= \gamma_1 + \gamma_2\\
    \label{eq:gamma_a}
    \gamma_a &= \frac{\gamma_1\gamma_2}{2\nu_c}\left[1+\left(\frac{\Delta\omega_1 - \Delta\omega_2}{\gamma_1 + \gamma_2}\right)^2\right]\\
    \label{eq:Delta_a}
    \Delta \omega_a &= \frac{\Delta\omega_1 \gamma_2 + \Delta\omega_2 \gamma_1}{\gamma_1 + \gamma_2}\\
    \label{eq:Delta_c}
    \Delta\omega_c &= \frac{\Delta\omega_1\gamma_1 + \Delta\omega_2 \gamma_2}{\gamma_1 + \gamma_2}
\end{align}

\noindent with $\nu_c = \frac{c}{2L_c}$ the free spectral range (FSR) of the c-cavity, $\gamma_i = \frac{cT_i}{4L_i}$ and $\Delta\omega_i$ the linewidth and detuning of cavity labeled $i$, where $T_i$ is the input mirror transmissivity of the $i$-th cavity.

We recall that the values for $\gamma_1, \gamma_2, \Delta\omega_1$ and $\Delta\omega_2$ are fixed for a given configuration of the interferometer \cite{Harms_2003}, in the lossless case. This means that a change in the cavity lengths $L_a, L_c$ of the 2FC setup should be accompanied by an equal change in the input mirrors' transmissions $T_1, T_2$ to keep $\gamma_1, \gamma_2$ constant. From \cref{eq:gamma_c}, this translates into $T_c \propto L_c$ while \cref{eq:gamma_a} implies that $T_a\propto L_aL_c \propto L_c^2$, where that last scaling law reasonably assumes that $L_a\propto L_c$.

\cref{table:params_general} provides the parameters which will be used in the rest of this paper, generally similar to previous studies \cite{Zhang24}. We briefly comment on the chosen value of 30 ppm for round trip loss per cavity $\Lambda$. They have been estimated according to the empirical scaling law: 
\begin{equation}
\Lambda = 30 \text{ ppm} \times  \left (\frac{L}{300 \text{ m}}\right)^{0.3} \left (\frac{1064 \text{ nm}}{\lambda} \right )^2
\end{equation}

\noindent where the values of 30 ppm has been found by simulations in \cite{zhao_24} for a 300-m cavity illuminated with a 1064 nm laser. For the scaling law with the length of the cavity we use the result of \cite{Isogai:13}. The scaling with the laser wavelength is given by the physics of the scattering \cite{PhysRevD.93.082004}. This can be considered a conservative estimation, since we expect future advances in the mirror and coating techniques. 

\begin{table}[h!]
\noindent
\begin{tabular}{l c r}
 \hline
 \hline
 Parameter & Physical meaning & Value \\ [0.5ex] 
 \hline
$\lambda$ & Laser wavelength & 1550 nm\\
$I_\text{arm}$ & Power in the arms & 18 kW\\
$T_\text{arm}$ & ITM transmission & 0.007 \\
$L_\text{arm}$ & Interferometer arm length & 10 km \\
$m_\text{ETM}$ & Mass of test-mass & 211.3 kg\\
$T_\text{SRM}$ & SRM transmission & 0.2 \\
$L_\text{SEC}$ & SEC length & 100 m\\
$\phi_\text{SEC}$ & SEC detuning & $\frac{\pi - 0.75}{2}$ rad\\
$\Lambda_\text{in}$ & Injection Faraday loss & $0.01$\\
$\Lambda_\text{arm}$ & Arm round-trip loss & $45$ ppm \\
$\Lambda_\text{SR}$ & SEC round trip loss & $1000$ ppm\\
\hline
$\theta_\text{HD}$ & Homodyne angle & $- 0.27$  rad\\
$\delta \theta_\text{HD}$ & HD RMS phase noise & 10 mrad (typ.)\\
$\Lambda_\text{out}$ & Readout loss & 0.03 \\
\hline
$\gamma_1$ & 1st FC linewidth & $2\pi \times (4.26)$ rad/s\\
$\gamma_2$ & 2nd FC linewidth & $2\pi \times (1.65)$ rad/s\\
$\Delta \omega_1$ & 1st FC detuning & $2\pi \times 19.51$ rad/s\\
$\Delta \omega_2$ & 2nd FC detuning & $2\pi \times (-7.65)$ rad/s\\
$\Lambda$ & Round Trip Loss per FC& 30 ppm \\
$\Lambda_\text{sub}$ & Substrate Loss (CFC) & 20 ppm \\
$\delta L$ & FC RMS length noise & 1 pm \\
$\Upsilon_I$ & Input mode mismatch & 4\%\\
$\Upsilon_O$ & Output mode mismatch & 3\%\\
$\Upsilon_{12}$ & Mode mismatch FC1-FC2 & 1\%\\
$\Upsilon_{a}$ & Internal mismatch (CFC) & $\sim 0$\%\\
$\Lambda_{FI}$ & FI loss (double pass) & 1\%\\
\hline
\hline
$L_{\text{FC1,2}}$ & Length of each FC & $5$ km\\
$L_{\text{CFC}}$ & Total length of CFC & $10$ km\\
$T_{\text{1}}$ & 1st FC input transmission & $6.9\cdot10^{-4}$\\
$T_{\text{2}}$ & 2nd FC input  transmission & $1.8\cdot10^{-3}$\\
$T_{\text{a}}$ & CFC middle transmission & $6.75\cdot10^{-6}$\\
$T_{\text{c}}$ & CFC input transmission & $2.47\cdot10^{-3}$\\
\hline
\hline
$L_{\text{FC1,2}}$ & Length of each FC & $1$ km\\
$L_{\text{CFC}}$ & Total length of  CFC & $2$ km\\
$T_{\text{1}}$ & 1st FC input  transmission & $1.4\cdot10^{-4}$\\
$T_{\text{2}}$ & 2nd FC input  transmission & $3.6\cdot10^{-4}$\\
$T_{\text{a}}$ & CFC middle transmission & $2.7\cdot10^{-7}$\\
$T_{\text{c}}$ & CFC input transmission & $4.95\cdot10^{-4}$\\

\hline
\hline
\end{tabular}
\caption{Parameters used in this study. For comparison, we show the parameters of 2FC and CFC if the total length is either 2 or 10 km. The round trip losses $\Lambda_\text{arm}, \Lambda_\text{SR}, \Lambda$ are computed per cavity of the interferometer (Arm or Signal Extraction Cavity) and frequency-dependent systems (2FC or CFC); the substrate loss $\Lambda_\text{sub}$ for the CFC case is accounted separately.}
\label{table:params_general}
\end{table}

\subsection{The middle mirror in CFC}
The middle mirror is a critical component of the CFC configuration.
We first comment on concerns raised by the unusually small value for $T_a$. First, note that while its value in the $L_\text{CFC} = 2$ km case ($T_a < 0.3$ ppm) is indeed much lower than what is implemented in the current generation of GW detectors, the $L_\text{CFC} = 10$ km case ($T_a = 6.75$ ppm) is comparable to the transmission of the end mirror of the currently installed filter cavity in Virgo (see \cite{AdVp_TDR}
), thus technically already achieved for wavelengths in the infrared spectrum. Therefore, in the rest of this paper, unless explicitly specified otherwise, we will work with $L_\text{CFC} = 10$ km, which seemingly is within technological reach.

Interestingly, if one considers the a-cavity as a single cavity -- that is, the input mirror is $T_a$ -- then it may seem that such a cavity is loss-dominated ($\Lambda = 30 \text{ ppm} \gg T_a $), and that the finesse would be of the order of $10^5$. However, since the a-cavity is embedded in a coupled cavity system, the correct qualitative picture is to consider that the first two mirrors ($T_c$ and $T_a$) form a cavity, which is equivalent to an effective mirror with variable transmission $T_\text{eff}$ given as a function of the sideband frequency $\Omega$ by
\begin{equation}
    T_\text{eff}(\Omega) = \frac{T_aT_c (1-\Lambda)}{1+ r^2 - 2 r\, \cos\left(\frac{2L(\Omega - \Delta\omega_c)}{c}\right)}
\end{equation}

\noindent with $r = \sqrt{R_aR_c(1-\Lambda)}$. Since $\Delta \omega_c = 12$ Hz and we are interested in sideband frequencies in the range of the expected GW signals $\Omega \in 2\pi\times [-30 \text{ Hz}, 30 \text{ Hz}] $, we have that $\Omega-\Delta\omega_c \in 2\pi\times [-42 \text{ Hz}, 18 \text{ Hz}] $. Plugging this into the previous formula, we find that $T_\text{eff} \in [210 \text{ ppm}, \ 10600 \text{ ppm}]$. This means that the effective second cavity is never loss-dominated in the range of frequencies of interest. The associated finesse range is $\mathcal{F}\in[6\cdot 10^2, \ 3\cdot 10^4]$ 
, whose upper value's order of magnitude is comparable to that of current squeezing filter cavities in LIGO \& Virgo  ($\mathcal{F} \sim 10^4$, see \cite{aligo_FDS,virgo_FDS}).

Technical limitations in the design of low-transmissivity mirrors may lead to small deviations from the theoretical optimal value $T_{\text{a}}$, as computed in \cref{table:params_general}, which in turn imply important reductions of the GW detector's sensitivity. These deviations can be directly or indirectly compensated, respectively through thermal actuators on the mirror or detuning offsets on the cavities' resonant frequencies.

Thermal controls enable direct tunability of $T_a$ in the following way: the two planar surfaces (Anti-Reflective and Highly-Reflective) of the middle mirror can be made parallel to each other to form an etalon \cite{Hild_2009}. Then, a controlled variation in the temperature of the mirror substrate changes the substrate index of refraction (thermo-refractive effect) as well as the physical length of the substrate (thermo-expansion), thus modifying the optical path length inside of the mirror \cite{galaxies8040080}. This can usually be done with a precision that should help mitigate imperfections in the initial value of $T_a$. Denoting $R_\text{AR}$ the reflectivity of the AR surface, we find that the transmission as a function of substrate temperature offset $\Delta \theta$ from room temperature has the form $T_a(\Delta \theta) \simeq  T_{a} [1+ 2 \sqrt{R_\text{AR}} \cos(2\pi \Delta \theta/\Delta\theta_T)]$, where $\Delta \theta_T = 1.49$ K for typical values in Fused Silica substrates. This means that a choice of $R_\text{AR} = 0.01$ allows to compensate a $20\%$ deviation of the value $T_a$ by tuning the mirror temperate. With a typical temperature precision of $\Delta\theta_\text{min} \sim 10$ mK, this scheme can control $T_a$ down to a relative error of $|\Delta T_a/T_a|\sim 1\%$, which is more than an order of magnitude improvement and ensures that the residual degradation of the GW detector sensitivity is negligible.

\begin{figure}[!h]
    \centering
    \includegraphics[width=\linewidth]{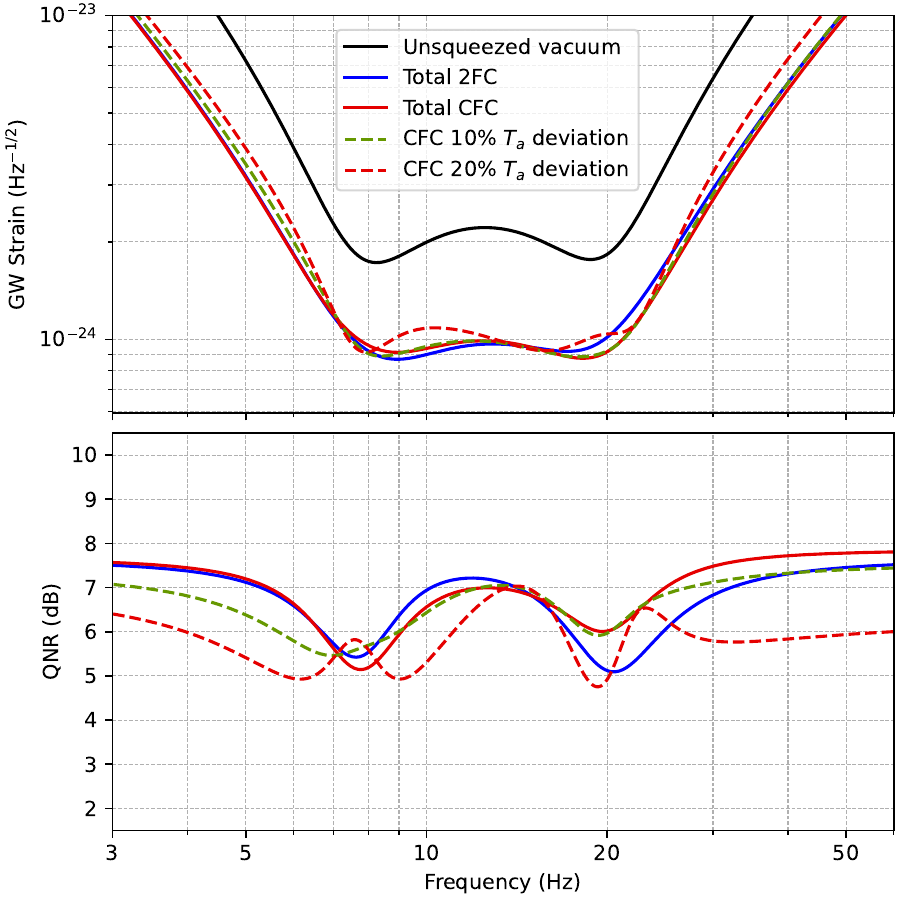}
    \caption{Squeezing degradation for the CFC caused by 10\% (dashed green) or 20\% (dashed red) deviations of the middle mirror's transmission $T_{\text{a}}$, compared to the CFC model with the optimal $T_{\text{a}}$ (solid red) and the two FC case (blue). The sensitivity without squeezing (solid black) is plotted for reference. The 10\% and 20\% $T_{\text{a}}$ deviations are being partly compensated by adjusting the detunings of the CFC sub-cavities and the injected squeezing angle. Losses are detailed in \cref{table:params_general} and \cref{sec:full_degradation}. Top:  Noise spectral density comparison for a single L-shaped interferometer. Bottom: Quantum enhancement in dB.}
    \label{fig:CFC MM compensation det and sqz angle}
\end{figure}

If thermal controls are unavailable, it is still possible to partially recover the detector sensitivity by tuning other parameters of the coupled filter cavity. However, in a realistic scenario, after the construction of the coupled filter cavity, most of the parameters cannot be adjusted. Thus, deviations in $T_{\text{a}}$ can be partially compensated by tuning only three parameters: the input squeezing angle into the CFC cavities, and the two detunings $\Delta\omega_a$ and $\Delta\omega_c$ of the sub-cavities \cite{Whittle_2020}. In practice, these detunings are controlled by locking the cavity on resonance with an auxiliary laser, which is frequency-locked at an offset from the main laser. By precisely controlling this offset, one can arbitrarily adjust the resonance conditions of the cavity and thus its frequency detuning from the squeezed field. \cref{fig:CFC MM compensation det and sqz angle} compares quantum noise reduction for 2FC with the one in the CFC case with a deviation in $T_{\text{a}}$ of 10\% and 20\% from its nominal value, after such partial compensation. We considered all loss sources of \cref{table:params_general} and we have taken the total cavity length to be $10$ km.  Therefore, by going from 20\% deviation (dashed red), which is the currently achievable tolerance on such a low absolute value of $T_a$ \cite{Fresnel2025}, to 10\% (dashed green), we can gain up to 1.5 dB of quantum enhancement above 30 Hz. This underlines the importance of reaching stringent tolerances on $T_{\text{a}}$. 

\section{Comparison of the effect of different squeezing degradation sources}\label{sec:comp}
In this section, we provide separate comparisons of the CFC and 2FC systems to understand the contributions of the different degradation mechanisms (optical loss $\Lambda$, cavity locking precision $\delta L$, mode mismatch $(\Upsilon, \psi_\text{mm})$). We start with a default model for both 2FC and CFC which corresponds to their lossless configurations, and study each one of these three degradation mechanisms separately.

The general form of the quantum noise enhancement (input squeezing parameter $r$) measured at the output of a lossy passive system has the form \cite{LIGO_quantum_response}:
\begin{equation}
\label{eq:SQZ_measurement_general}
\begin{split}
    \bar{S}[\Omega] &= \eta[\Omega]  \{[(1-\Xi[\Omega] ) e^{-2r} + \Xi[\Omega] e^{2r}] \cos^2(\Delta\theta_D[\Omega]) \\
    & \quad \qquad  + [(1-\Xi[\Omega] ) e^{2r} + \Xi[\Omega] e^{-2r}] \sin^2(\Delta\theta_D[\Omega]) \}\\
    & \quad + 1-\eta[\Omega]
 \end{split}
 \end{equation}

\noindent where $\eta[\Omega]$, $\Xi[\Omega]$ and $\Delta\theta_D[\Omega]$ are the three frequency-dependent figures of merit that characterize the degradation, namely the efficiency, the dephasing and the misphasing. Their mathematical definitions are provided in \cref{app:notations}. Intuitively, 
\begin{itemize}
    \item $\eta[\Omega]$ characterizes how much input squeezing is lost to vacuum;
    \item $\Xi[\Omega]$, akin to phase noise, characterizes how much sideband imbalance is induced by the system, thus how much antisqueezing is effectively coupled to squeezing through a phenomenon known as "complex squeezing" \cite{ding2024};
    \item $\Delta\theta_D[\Omega]$ encapsulates the difference of the rotation of the squeezing ellipse between the realistic and the perfect (lossless) system.
\end{itemize}

\subsection{Optical Losses}\label{sec:optical_loss}
\begin{figure}
    \centering
    \includegraphics[width=\linewidth]{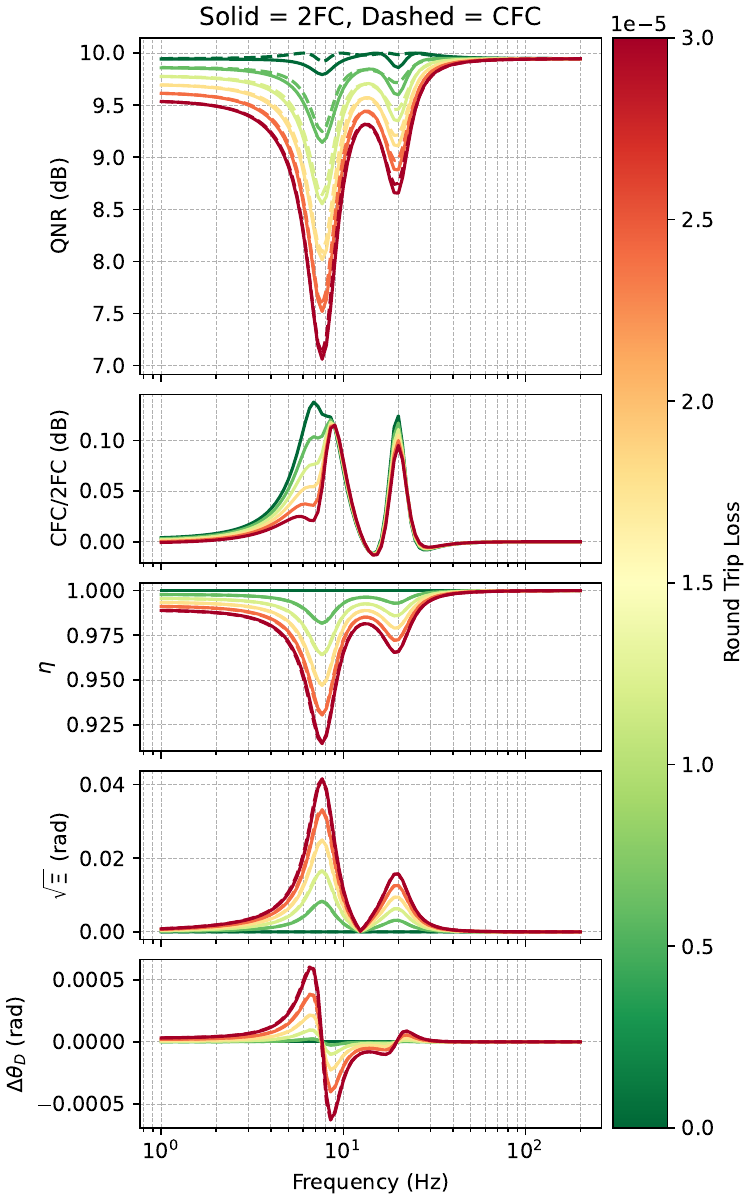}
    \caption{Squeezing degradation for Round Trip Loss per cavity $\Lambda$ for 2FC (solid lines) and CFC (dashed lines), for a total length $L_1 + L_2 = L_a + L_c =  5 \text{km} + 5 \text{km} = 10$ km. From top to bottom: Quantum noise reduction (for 10 dB of input squeezing); Relative quantum noise reduction between CFC and 2FC (a positive value means that CFC performs better than 2FC); the squeezing efficiency $\eta$, the dephasing $\sqrt{\Xi}$ and the misphasing $\Delta \theta_D$.}
    \label{fig:RTL_comparison}
\end{figure}

\cref{fig:RTL_comparison} compares the squeezing degradation when Round Trip Losses $\Lambda$ are added in the CFC and 2FC models (which can be physically interpreted as absorption loss on each mirror), for values of $\Lambda$ equal for all mirrors and ranging from 0 to 30 ppm per cavity. In the case of CFC, this excludes additional substrate loss due to passing through the middle mirror to go from one sub-cavity to the other. The substrate loss are considered separately in the full degradation budget.

Remarkably, we see that the degradation curves between CFC and 2FC are essentially identical for a given $\Lambda$. This was also noted in \cite{Jones_2020} but no proof was given. This is proved in \cref{app:proof_scaling_laws}, where we we show that if CFC and 2FC have the same round trip losses (RTL), then they have the same response at the lowest order. We also see that the squeezing efficiency and dephasing reach extrema at the resonance frequencies $|\Delta\omega_{1,2}|$, while misphasing is comparatively negligible ($|\Delta \theta_D| \ll \sqrt{\Xi}$). 

Finally, let us note that, because the middle mirror in CFC is being traversed by the optical field going from the $c-$cavity to the $a-$cavity, substrate loss may further degrade squeezing. It turns out (\cref{app:sub_loss}) that, for current substrate loss values, this extra degradation is always inferior to 0.5 dB of squeezing thus is not significantly worsening the performance of CFC.

\subsection{Mode mismatch}

Mode mismatching between cavities and free-space beams has been hypothesized as the main contributor of unknown loss in current gravitational wave detectors. Its modeling is in general more involved than loss as the power of the incident beam is unitarily redistributed among higher order modes \cite{LIGO_quantum_response} and thus requires several parameters to characterize which depend on the overlap between the incident and the cavity modes. 

Importantly, mode coupling can manifest in two distinct effective ways: so-called quadratic mismatch, due to waist position/size mismatch or other large-scale defects of the optics; or involve higher order aberrations, typically due to point defects and high-order scattering, see \cite{Kuns2025}. The former is modeled using a unitary basis change $\boldsymbol{U}$, while the latter usually corresponds to modes which are stopped by baffles and thus can simply be modeled using an absorption coefficient $\Lambda$. In the following, we will focus our discussion on the quadratic type, that is, we assume that input, intra-cavity, and output mismatches are all modeled by unitary matrices $\boldsymbol{U}$.

We make the general assumption that mode mismatch occurs when a beam goes from free space into a cavity and vice-versa, because of limited actuation in the mode-matching telescopes and in the possible low-order aberrations that they induce. However, for coupled cavities, the mode mismatch $\Upsilon_a$ between the two subcavities only depends on the design of the cavity mirrors, which can be made with much lower imperfections and thus this mode mismatch can made negligible (see \cref{app:MMCFC} for a quantitative derivation). 

We work in several eigenbases when propagating the modes throughout the cavities. We consider the transverse electromagnetic bases of the input field (index $I$), the eigenbasis of the first cavity (index $1$ for 2FC, $c$ for CFC), the eigenbasis of the second cavity (index $2$ for 2FC, $a$ for CFC), the eigenbasis of the "output" field which exits the FDS system and gets injected into the interferometer (index $O$). 
In order to have tractable insights in our model, we only consider a single higher order mode that encapsulates all the mode mismatch for each basis. We regroup these two modes in a vector $\boldsymbol{a}[\Omega] = \begin{bmatrix}
    a_1[\Omega]\\
    a_2[\Omega]
\end{bmatrix}$ where the first element is the fundamental gaussian mode, while the second is the higher order mode, evaluated at the sideband frequency $\Omega$ from the carrier (see \cref{app:notations} for details). Mode mismatch is characterized by two parameters, $\Upsilon\in[0, 1]$ the amplitude of the mismatch ($\Upsilon = 0$ corresponds to no mismatch), and $\psi_\text{mm}$  the phase of the mode mismatch. For example, if we are interested in the matching between a free-space beam and a cavity eigenmode, $\psi_\text{mm} = 0$ means that all the mismatch comes from beam waist size,  while $\psi_\text{mm} = \pi/2$ means that all the mismatch comes from beam waist longitudinal position \cite{Anderson:84}. An intermediary value of $\psi_\text{mm}$ corresponds to a linear combination of waist size and position mismatch.

\begin{figure}
    \centering
    \includegraphics[width=\linewidth]{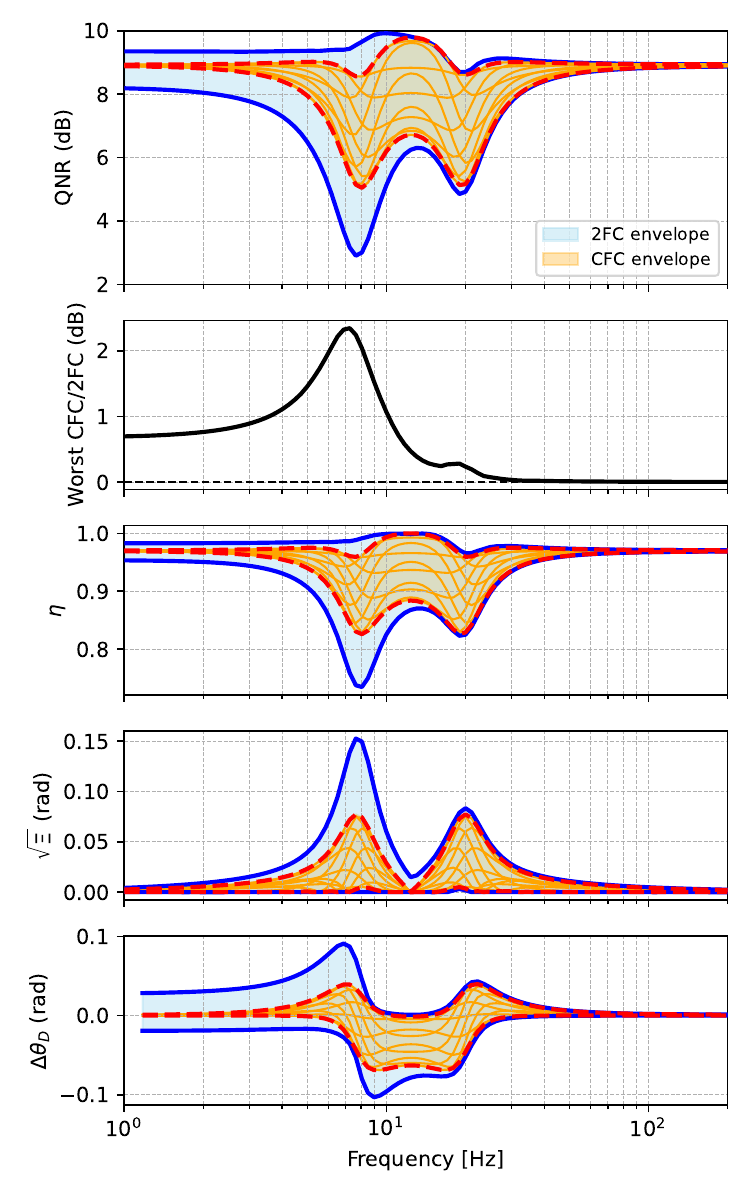}
    \caption{Squeezing degradation from quadratic mode mismatch in 2FC (blue, solid envelope) and CFC (red, dashed envelope), for a total length $L_1 + L_2 = L_a + L_c =  5 \text{km} + 5 \text{km} = 10$ km. The blue and orange regions are determined by calculating squeezing curves for the values of $\Upsilon_I$, $\Upsilon_O$, $\Upsilon_{12}$ and $\Upsilon_a$ from \cref{table:params_general}, and sweeping the mode mismatch phases ($\psi_{mmO} - \psi_{mmI}$) and ($\psi_{mm12} - \psi_{mmI}$). The inter-cavity mode mismatch 2FC is $\Upsilon_{12} = 0.01$. Some examples of CFC squeezing curves are plotted in orange. From top to bottom: quantum noise reduction with mismatch; ratio of the lower envelopes of CFC to 2FC; efficiency; dephasing; misphasing.}
    \label{fig:CFC_MM_envelope}
\end{figure}

Under the quadratic mismatch model, we see in \cref{fig:CFC_MM_envelope} that squeezing degradation presents a wideband feature characterized by the value of QNR at $\Omega \to +\infty$, and, in the worst case (lower envelopes), peaks of degradation close to the detuning frequencies $\Delta \omega_{1,2}$. To interpret these features, let us compute approximate forms for the squeezing efficiency $\eta$ and dephasing $\Xi$ at these frequencies. 

We see from the numerical values of \cref{table:params_general} that $\Delta \omega_1^2 \gg \gamma_1^2$ and similarly $\Delta \omega_2^2 \gg \gamma_2^2$, while $\Delta\omega_1-\Delta \omega_2 > \gamma_1, \gamma_2 $. This means that we can reasonably treat both cavities separately. When the higher order mode is not resonant inside of a single cavity, its response to the squeezed mode is shown to be \cite{LIGO_quantum_response}
\begin{equation}\label{eq:F_1FC}
    F_{1FC}[\Omega] = \sqrt{1-\Upsilon_O}\frac{\ii (\Omega+ \Delta\omega) - \alpha\gamma}{\ii(\Omega +\Delta\omega) + \gamma}
\end{equation}
\noindent where $\gamma, \Delta\omega$ are the bandwidth and detuning of the cavity, $\alpha = 1-2\Upsilon_I + 2\beta \sqrt{\Upsilon_I\Upsilon_O}\, e^{\ii\psi_R}$ with $\beta = \sqrt{\frac{1-\Upsilon_I}{1-\Upsilon_O}}\simeq 1$ and $\psi_R = \psi_O + \psi_G - \psi_I$. The figures of merit are (see \cref{eq:eta,eq:Xi})
\begin{align}
    \eta[\Omega] &= \frac{|F_{1FC}[\Omega]|^2 + |F_{1FC}[-\Omega]|^2}{2}\\
    \Xi[\Omega] &= \frac{\left(|F_{1FC}[\Omega]| - |F_{1FC}[-\Omega]|\right)^2}{4\eta[\Omega]}
\end{align}

At $\Omega \to \infty$, we find 
\begin{align}
    \eta[\Omega\to\infty] &\simeq 1-\Upsilon_O \\\Xi[\Omega\to\infty]  &= 0
\end{align}
which means that for CFC, which can be mapped to two single filter cavities as seen above, the high-frequency degradation is entirely determined by the output mismatch $\Upsilon_O$ and only shows up as loss. Numerically, for 10 dB of injected squeezing, this would correspond to a QNR of $\sim 9$ dB, consistent with the value simulated numerically. At the frequencies of the detunings, we have
\begin{align}
F_{1FC}[-\Delta\omega_{1,2}]&= -\alpha\sqrt{1-\Upsilon_O}\\F_{1FC}[\Delta\omega_{1,2}]&= \sqrt{1-\Upsilon_O}.
\end{align}
Thus \begin{align}
\eta[\Delta\omega_{1,2}] &= (1-\Upsilon_O)\frac{1+|\alpha|^2}{2} \\\Xi[\Delta\omega_{1,2}] &= \frac{1}{2} - \frac{|\alpha|}{1+|\alpha|^2}
\end{align} while the misphasing is given by $\Delta \theta[-\Delta\omega_{1,2}] = \arg(\alpha)$. The lowest efficiency and highest dephasing are obtained when $\psi_R = \pi$, that is, the higher order mode constructively interferes to degrade squeezing. Using the values in \cref{table:params_general}, we estimate a QNR of $\sim 5$ dB, consistent with the plotted minima at around $\Delta\omega_{1,2}$. Thus, even for relatively low mismatch amplitude, squeezing is significantly degraded.

For 2FC, because the internal mode mismatch $\Upsilon_{12}$ is not zero in general, the expression of the response is more involved. The matrix form is:
\begin{equation}\label{eq:F_2FC}
\begin{split}
    \boldsymbol{F}_{2FC} &= \boldsymbol{U}(\Upsilon_O, \psi_{mmO})\boldsymbol{U}(\Upsilon_I, \psi_{mmI})^\dagger\, \boldsymbol{U}(\Upsilon_{12}, \psi_{mm12})^\dagger \\
    &\qquad \times \boldsymbol{F}_2 \, \boldsymbol{U}(\Upsilon_{12}, \psi_{mm12}) \boldsymbol{F}_1\boldsymbol{U}(\Upsilon_I, \psi_{mmI})
\end{split}
\end{equation}

\noindent where $\boldsymbol{F}_1 = \begin{bmatrix}
    r_1 & 0\\
    0 & 1
\end{bmatrix}$ with $r_1 = -\frac{ \gamma_1 - \ii (\Omega+\Delta\omega_1)}{\gamma_1 + \ii (\Omega+\Delta\omega_1)}$, and analogously for $\boldsymbol{F}_2$. We are only interested in the upper left matrix element of $\boldsymbol{F}_{2FC}$; however its general form is cumbersome. We can nevertheless determine the squeezing degradation at high frequencies $\Omega\gg \Delta\omega_1, \Delta\omega_2$ as the transfer matrices simplify to $\boldsymbol{F}_1 = \boldsymbol{F}_2 = \boldsymbol{1}$ so $F^{(11)}_{2FC}[\pm \Omega] \underset{\Omega \to \infty}{=} \sqrt{\Upsilon_O}$, similarly to the CFC behavior.

To study the degradation in 2FC close to the detuning frequencies, notice in the formula \cref{eq:F_2FC}, the first cavity ``sees"  a mismatch of $\boldsymbol{U}(\Upsilon_I, \psi_{mmI})$, while the second cavity sees a mismatch of $\boldsymbol{U}(\Upsilon_{12}, \psi_{mm12})\boldsymbol{U}(\Upsilon_I, \psi_{mmI})$. This means that, outside of the resonance of the second cavity ($r_2 \simeq 1$), the response $\boldsymbol{F}_{2FC}$ is independent of $\Upsilon_{12}$ and its upper left element can be approximated by \cref{eq:F_1FC}. Thus at the resonance of the first cavity, we expect similar degradation than the CFC case (QNR of 5 dB for the values in \cref{table:params_general}). On the contrary, outside of the resonance of the first cavity ($r_1 \simeq 1$), $F_{2FC}$ can be approximated by $F_{1FC}$, but provided that one makes the substitution $\boldsymbol{U}(\Upsilon_I, \psi_{mmI}) \to \boldsymbol{U}(\Upsilon_{12}, \psi_{mm12})\boldsymbol{U}(\Upsilon_I, \psi_{mmI})$. We then redefine $\alpha = 1 - 2\Upsilon' + 2\beta' \sqrt{\Upsilon'\Upsilon_O} e^{\ii \psi'}$ where $\beta' = \sqrt{\frac{1-\Upsilon'}{1-\Upsilon_O}}$ and the expressions of $\Upsilon'$ and $\psi'$ are given in \cref{eq:Upsilon_prime} of \cref{app:add_mm}. Importantly, the maximum amount of mismatching thus obtained is larger than the sum of the mismatches: $ \Upsilon'_\text{max} = 0.088> \Upsilon_I + \Upsilon_{12} = 0.05$, which means that the degradation at the second cavity's resonance can be larger than the first's, as shown in the figure. Numerically, at $\Delta\omega_2$, we find, by replacing $\alpha$ with $\alpha'$ in the formulas for 1FC, that in the worst case ($\Upsilon' = \Upsilon'_\text{max}$ and $\psi' = \pi$), $\eta[\Delta\omega_{2}] = 0.74$ and $\Xi[\Delta\omega_2] = 0.025$, implying that QNR $\sim 3$ dB, as shown in \cref{fig:CFC_MM_envelope}. This highlights that mode mismatch does not affect both cavities similarly, as the second cavity may experience larger degradation around its frequencies of resonance. This means that, in order to favor low frequencies, it may be beneficial to place the cavity with the lowest detuning first, opposite to what is shown in \cref{fig:CFC_MM_envelope}.

\medskip
\subsection{Filter cavity length fluctuations}
\begin{figure}
    \centering
    \includegraphics[width=\linewidth]{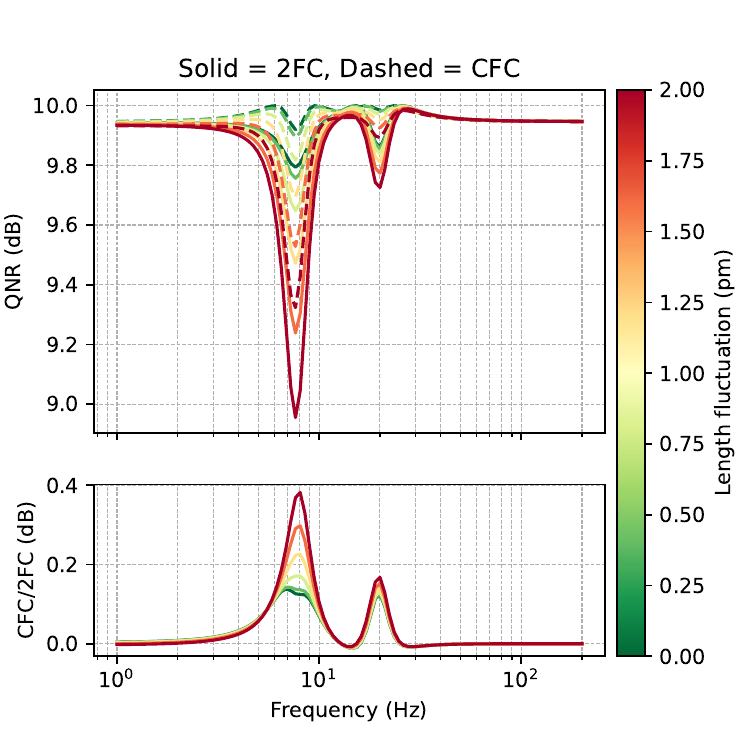}
    \caption{Squeezing degradation for length fluctuation $\delta L$ for 2FC (solid lines) and CFC (dashed lines), for a total length $L_1 + L_2 = L_a + L_c =  5 \text{km} + 5 \text{km} = 10$ km.}
    \label{fig:length_noise_comparison}
\end{figure}

The locking imprecision on the filter cavities induce a length fluctuation $\delta L$ which converts into phase noise for the squeezed field. Using the model developed in \cref{app:phase_noise}, we plot its effect on squeezing in \cref{fig:length_noise_comparison}. The input squeezing level is 10 dB. We see that overall, cavity length fluctuations of less than 2 pm do not significantly degrade squeezing, neither for CFC nor 2FC, with a marginally better performance of CFC. Importantly, we have assumed that all the cavity lengths ($L_1$ and $L_2$ for the 2FC; $L_c$ and $L_a$ for the CFC) have the same fluctuation RMS $\delta L$. In practice, this will depend on the actual control scheme used to stabilize each of these cavities. While we defer the concrete technical control scheme to future work, we note that the range of values chosen for this length noise is realistic, as an FDS cavity  controlled down to $\delta L \sim 1$ pm was achieved in the current generation of GW detectors \cite{aligo_FDS,virgo_FDS}.

\subsection{Full degradation budget comparison}
\label{sec:full_degradation}

\begin{figure*}
    \centering    \includegraphics[width=0.8\linewidth]{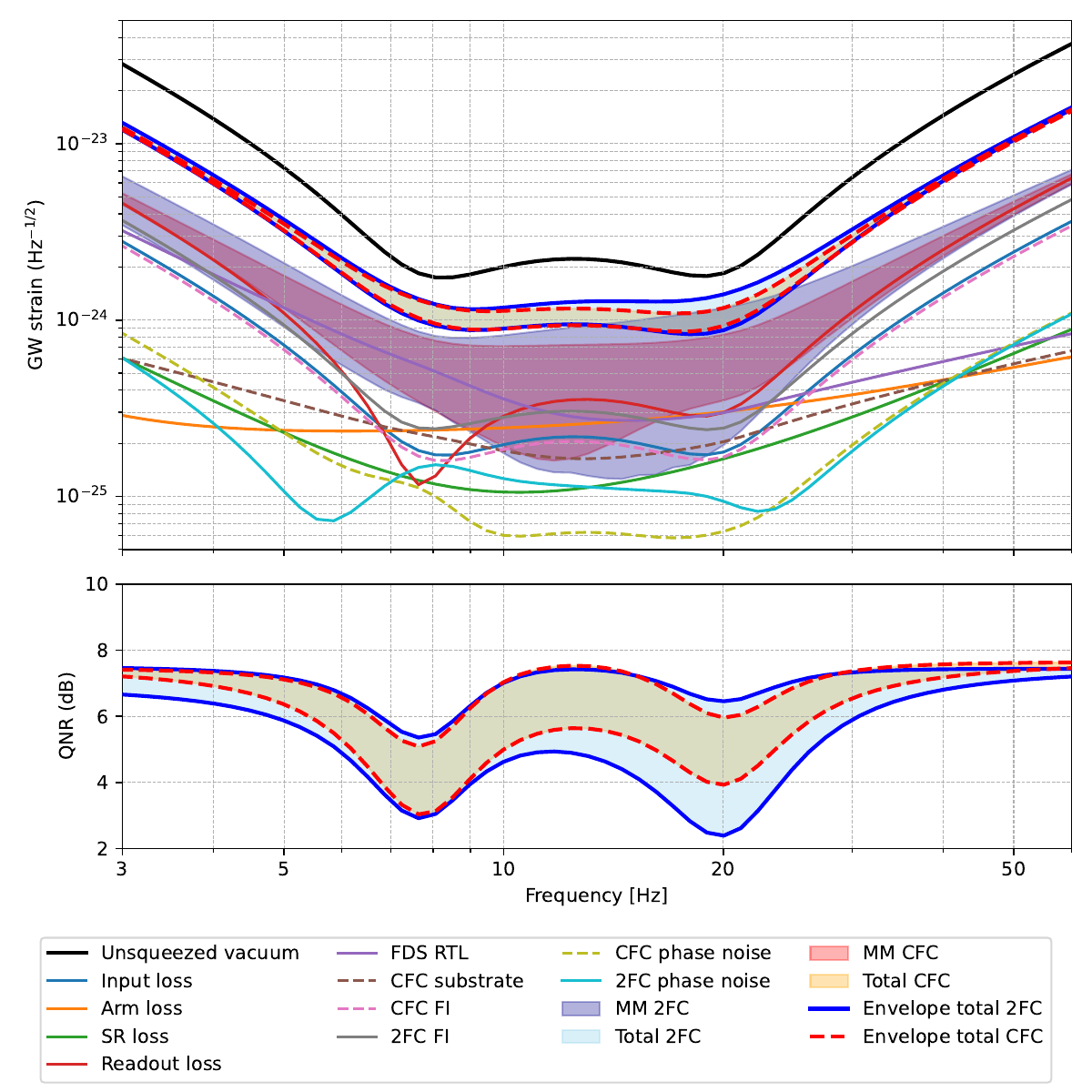}
    \caption{Comparison of quantum noise degradation in each ET-LF interferometer, between CFC (red dashed curve) and 2FC (blue solid curve), using the full budget of \cref{table:params_general}. We have chosen the arms angle to be 90$^o$ for simplicity, so that the strain is equal to the differential arm length.}
    \label{fig:CFC_2FC_full}
\end{figure*}

Having individually studied all three loss contributions, we now determine the full picture of loss degradation in a 2FC and CFC. The response is illustrated in \cref{fig:CFC_2FC_full}. The non-linear addition of mode mismatches in 2FC, along with the extra optical losses of the Injection Faraday isolator, ultimately degrade the total squeezing level below that of the CFC scheme. We also include interferometer losses present due to the injection Faraday isolators (see \cref{fig:ITF}) \footnote{We retain a Faraday injection loss of $1\%$ while in the rest of the optical layout, each single-pass individual Faraday isolator incurs a loss of $0.5\%$. This is consistent because in practice, to greatly minimize backreflection, we actually require two Faraday isolators in the injection stage, so $2 \times 0.5 = 1\%$.}, in the arm cavities, in the Signal Extraction Cavity and at the readout (see \cref{table:params_general} for their values and \cref{app:ITF_response} for explicit derivation of the loss contributions). This study thus concludes that, all other parameters and degradation sources being equal, CFC overall performs similarly or slightly better than 2FC.

\section{Tuned single filter cavity configuration for Einstein Telescope}
\label{sec:applications}

\begin{figure}
    \centering
    \includegraphics[width=0.8\linewidth]{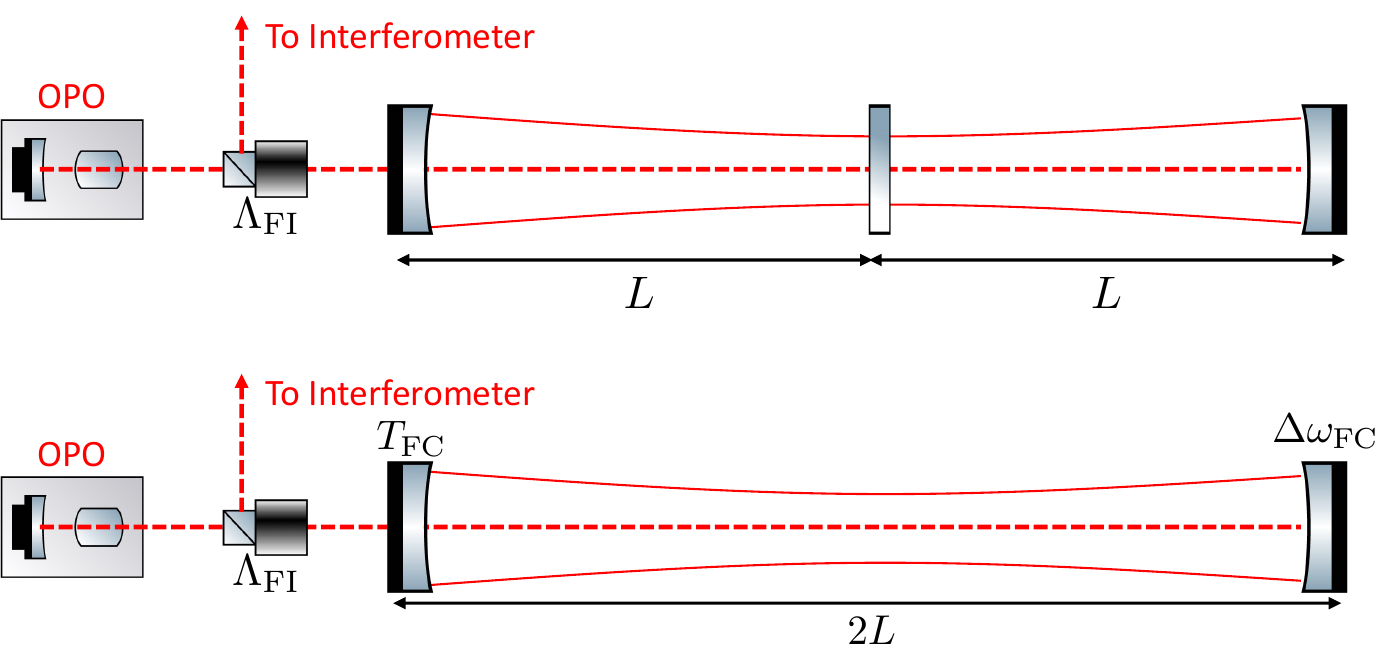}
    \caption{Comparison of CFC (top) and 1FC (bottom) schemes, respectively for detuned and tuned ET-LF. One swaps from CFC to 1FC by removing the middle mirror and changing the input mirror transmissivity $T_c \to T_\text{FC}$.}
    \label{fig:1FC}
\end{figure}

It has been repeatedly noted in the literature that the practical operation of a GW interferometer with a detuned signal extraction cavity poses significant technical challenges in the stabilization of the various control loops \cite{Ganapathy_2021,Ward2010length}. In this section, we propose an alternative to ET-LF that corresponds to a tuned configuration, along with a single filter cavity for optimal frequency-dependent squeezing (referred henceforth as the 1FC configuration). This would serve as either 1) an intermediary and easier-to-control configuration before upgrading to the full detuned setpoint or 2) a fallback solution, in case the detuned setpoint of ET-LF is shown to be technically difficult to stabilize. Importantly, we note that the 1FC configuration uses essentially the same infrastructure and excavation footprint as CFC, so that no significant overhead may be required in terms of hardware, other than changing the mirror transmissivities. The optical layout is shown in \cref{fig:1FC} and compared to CFC: the length of the 1FC cavity is $L_\text{1FC} = 2L = L_a + L_c$.

\begin{figure}
    \centering
    \includegraphics[width=\linewidth]{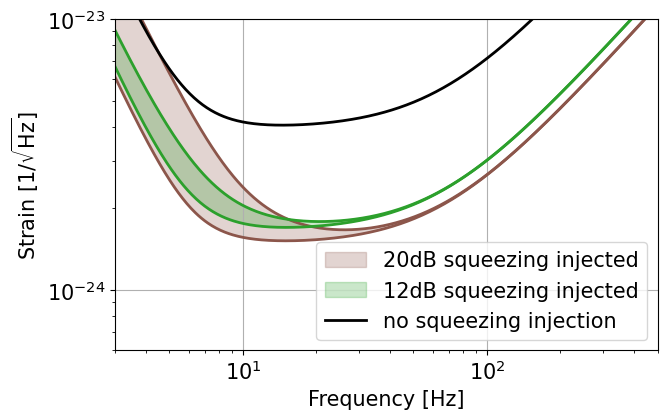}
    \caption{Quantum noise of ET-LF when the signal extraction cavity is tuned and a single 10 km filter cavity is used for frequency dependent squeezing. Different color curves show quantum noise with different level of input squeezing. The shaded area represents the variation of quantum noise due to the unconstrained mode mismatch phases. All parameters except the tuning of the signal extraction cavity $\phi_\text{SEC}$ are identical to \cref{table:params_general}.}
    \label{fig:ETLF_tuned_1220sqz}
\end{figure}

Previous studies of tuned ET-LF with 1FC, such as \cite{Jones_2020,ETlayout2024}, only considered quantum noise degradation due to round trip loss and did not assume equal footprint $L_\text{1FC} = L_a +L_c$ in their comparison to detuned ET-LF + CFC/2FC. By keeping the total 1FC/CFC/2FC lengths to be equal and also including dephasing loss induced by mode mismatch (same values as in \cref{table:params_general}), we find that an increase in injected squeezing level negatively impacts the total sensitivity, as the mode mismatch phases are in general unconstrained, as shown in \cref{fig:ETLF_tuned_1220sqz}, where an injected squeezing level of $~20$ dB (as initially studied in \cite{Zhang24}) may lead to worse performance for some choices of mismatch phases than lower levels (12 dB). This confirms that, even in the case of a single filter cavity and tuned ET-LF, mode mismatch remains a limitation for squeezing injection.

\begin{figure}
    \centering
    \includegraphics[width=\linewidth]{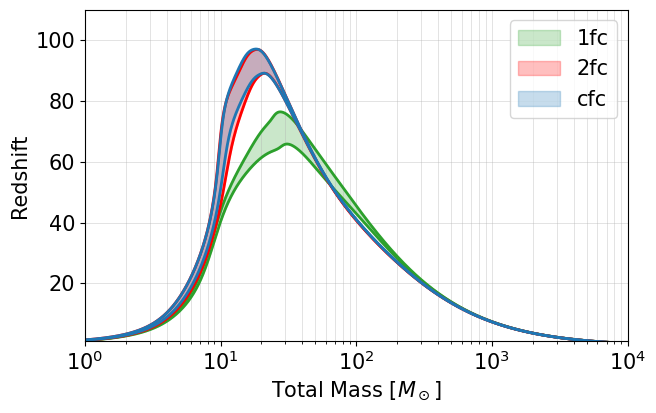}
    \caption{Astrophysical reach of ET when ET-LF is tuned or not detuned, assuming identical total length of CFC, 2FC and 1FC (10 km) and a geometry given by 60$^o$ interferometers in a triangular network. The shaded areas represent the possible redshifts due to the unconstrained mode mismatch angles for each of the configurations.}
    \label{fig:2fc1fc}
\end{figure}

We can further provide a comparison with the full noise budget, which includes contributions from quantum noise and other sources related to the interferometer such as residual gas noise, thermal noise, newtonian noise, seismic noise.
The quantum noise curves for CFC and 2FC are obtained using the method described in \cref{sec:comp}, then added to the other classical noises (see \cite{ETlayout2024}) and converted to redshift units. To maximize the astrophysical reach of tuned ET + 1FC, we optimize on the redshift and use as free parameters the signal recycling transmissivity $T_\text{SRM}$, the transmissivity of the 1FC's input mirror $T_\text{FC}$, {the injected squeezing level} $r_\text{tuned}$ as well as the detuning of 1FC $\Delta\omega_\text{FC}$. These are well-studied parameters to optimize over in previous and current generation interferometric detectors  \cite{purdue2002practical,kwee2014decoherence,Ganapathy_2021,Whittle_2020}. {We use a particle swarm optimization method \cite{venter2003particle} to maximize the astrophysical reach for total masses between 10 and 30 solar masses.} We also take into account the redshift contribution of the High-Frequency Einstein Telescope interferometer, as well as the triangular geometry of the full Einstein Telescope layout -- which simply amounts to an overall scaling factor \cite{Regimbau_2012}. The results for total length of $L_\text{1FC} = L_a + L_c = L_1 + L_2 = 10$ km are presented in Fig.\,\ref{fig:2fc1fc}, while the optimized parameters for 1FC are shown in \cref{table:params_tuned} \footnote{We recall that in this study, we consider a single filter cavity for frequency dependent squeezing; the optimized value of $T_\text{SRM}$ could be larger had we considered two filter cavities, since in the limit $T_\text{SRM} \to 1$ we obtain a power-recycled Michelson interferometer whose optimal squeezing response can only be obtained with two filter cavities \cite{klmtv}.}.

We see from \cref{fig:2fc1fc} that the CFC and 2FC both perform better (up to a factor of 2 in redshift) in the middle-mass range $[10 M_\odot, \ 30 M_\odot]$ due to their enhanced sensitivity in the corresponding frequency range of these mergers, while the tuned configuration offers comparable or better performance outside of this range. Nevertheless, we note that the values of mode mismatch have been taken to be conservative in this study, such that if these values are improved in practice, better performance for both tuned and detuned configurations may be expected. Furthermore, the tuned configuration still achieves orders-of-magnitude redshift improvement over current generation detectors such as Advanced LIGO (maximum redshift of the order of 1), and goes beyond the star formation epoch ($\sim 20$ redshift), which confirms this configuration's scientific interest.

\begin{table}[h!]
\centering
\begin{tabular}{l c r} 
 \hline
 \hline
 Parameter & Physical meaning & Value \\ [0.5ex] 
\hline
$T_{\text{FC}}$ & FC input mirror transmissivity & $0.37$ \%\\
$\Delta\omega_{\text{FC}}^{\mathrm{10km}}$ & FC detuning & $4.20$ Hz\\
$r_{\text{tuned}}$ & Injected squeezing & $12$ dB\\
$T_{\text{SRM}}$ & SR mirror transmissivity & $44$ \% \\
\hline
\end{tabular}
\caption{Optimized parameters for tuned ET-LF + 1FC. Apart from these values and the SEC detuning (set here to $\phi_\text{SEC} = \pi/2$ rad), the other interferometer parameters are identical to \cref{table:params_general}.}
\label{table:params_tuned}
\end{table}


\section{Conclusions and future steps}

Frequency dependent squeezing has been proved an effective techniques to reduce quantum noise in gravitational-wave detectors. In this work we have extended the previous studies about FDS requiring complex cavity systems to achieve the full rotation of the squeezing ellipse, and applied the results to ET-LF. We have seen that a 10-km CFC layout is achievable with standard mirror technologies, and deviations from nominal mirror parameters can be compensated. Importantly, the CFC configuration behaves equally or slightly better in terms of robustness with respect to squeezing degradation sources compared to the 2FC one. In particular, we have provided new theoretical insights in understanding the equivalency of CFC and 2FC layouts for round trip losses, as well as the propagation of cascaded mode mismatch on the degradation of squeezed light. For future work, special attention should nevertheless be given to controllability issues of the CFC scheme, expanding for example on \cite{control_CFC}. We also note that our insights on quantum degradation in multiple cavity schemes could be extended to hybrid EPR+1FC \cite{Zhang24} or quantum teleportation schemes \cite{Nishino_2024}, where the interferometer itself acts as an effective single or multi-cavity.

Armed with the detailed model of quantum noise, we have studied the tuned ET-LF + 1FC configuration -- similar to the layout of LIGO during the O4 run -- and compared it to the final detuned ET-LF + 2FC/CFC designs. This comparison is relevant because the footprint of both layouts are similar, a crucial point for an underground interferometer. The relative redshift gain of the detuned configuration compared to the tuned case incites us to consider a 2-step implementation for ET:  first attain the tuned configuration by removing the middle mirror of CFC, then, add the latter and detune the interferometer to achieve the designed ET-LF sensitivity. Such an approach, should its controllability be demonstrated, may simplify the overall optical scheme, and mitigate the effects of some squeezing degradation sources. Finally, we note that, up to some overall geometrical factor, the results of this paper are independent of the final geometry of the detector, thus remain relevant should either a triangular or a L-shape configuration be ultimately adopted. 

\FloatBarrier
\section*{Acknowledgments}
 We thank Michel Lequime, Myriam Zerrad, Anne Daumas, Michael Hartmann, Giancarlo Cella, Jérôme Degallaix, {Francesco Di Renzo, Mikhail Korobko}, Léon Vidal and Matthew Evans for the fruitful discussions and comments. This work has been supported by Quantum FRESCO (ANR-23-CE31-0004), by the Paris Region and by the EU Horizon 2020 Research and Innovation Program under the Marie Sklodowska-Curie Grant Agreement No. 101003460 "PROBES".

\appendix
\section{Formalism}\label{app:notations}
In this appendix, we describe the mathematical formalism used in the main sections of the paper.

\subsection{Notations, Conventions}
 A classical, monochromatic electric field at frequency $\omega >0$, which propagates along $(+x)$, is written in complex form $
        E(t,x) = E_0 \, \exp\left(-i\omega t + i k x\right)$ with $k = \frac{\omega}{c}$. In the presence of a cavity, we decompose the optical frequency $\omega = \omega_0 + \Omega = n\omega_\text{FSR} + \Delta \omega + \Omega$ where $\omega_0$ is the reference frequency of the bright coherent field (1550nm for ET LF), $\omega_\text{FSR}$ is the free spectral range of the cavity, $\Delta \omega$ the detuning of the cavity, $\Omega$ the sideband frequency with respect to $\omega$, and $n\in \mathbb{N}$.

\subsection{Matrix formalism in passive systems}
Let us consider two copropagating spatial modes $a_1$ and $a_2$ (for example TEM00 and TEM01) of the electromagnetic field, written in the ladder basis vector $\boldsymbol{a}[\Omega] = \begin{bmatrix}
    a_1[\Omega]\\
    a_2[\Omega]
\end{bmatrix}$. Then, the usual transformations are 

\begin{enumerate}
    \item Free propagation without loss:
    \begin{equation}
        \boldsymbol{L}(\Omega, \Delta L, \psi) = \exp\left(i\frac{\omega_0\Delta L+\Omega L_0}{c}\right)\begin{bmatrix} 1 & 0 \\
        0 & e^{i\psi}
        \end{bmatrix}
    \end{equation}
    \noindent where $\Delta L$ is the microscopic length detuning from the resonance, $L_0$ is the macroscopic length of the cavity on resonance, $\psi$ is the Gouy phase of the higher order mode $a_2$ relative to $a_1$. 
    
    \item 
    Power loss:
    \begin{equation}
        \boldsymbol{E} = \sqrt{1-\Lambda} \, \boldsymbol{1} 
    \end{equation}
    \noindent where $\Lambda$ is the power loss.

    \item Mirror reflection and transmission:
    \begin{equation}
        \boldsymbol{r} = \sqrt{1-T} \, \boldsymbol{1} \qquad \text{and} \qquad \boldsymbol{t} = \sqrt{T} \, \boldsymbol{1}
    \end{equation}
    \noindent where $T$ is the power transmissivity.
    
    \item Mode mismatch:
    \begin{equation}
        \boldsymbol{U}(\Upsilon, \psi_{mm}) = \begin{bmatrix}
            \sqrt{1-\Upsilon} & -\sqrt{\Upsilon} e^{\ii\psi_{mm}} \\
            \sqrt{\Upsilon} e^{-\ii\psi_{mm}} & \sqrt{1-\Upsilon}\\
        \end{bmatrix}
    \end{equation}

    \noindent where $\Upsilon$ is the amplitude of the mismatch, while $\psi_{mm}$ is the phase of the mismatch.
    
\end{enumerate}

\subsection{Quantum noise in passive systems}
Consider a system which is only composed of passive elements. Its classical transfer function on the signal mode is 
\begin{equation}
a^\text{out}[\Omega] = F[\Omega] a^\text{in}[\Omega]
\end{equation}

so in the two photon formalism, its matrix has the form
\begin{equation}
\boldsymbol{\mathfrak{a}}^\text{out}[\Omega] = \begin{bmatrix} 
F[\Omega] & 0\\
0 & F[-\Omega]^*
\end{bmatrix}
\boldsymbol{\mathfrak{a}}^\text{in}[\Omega] = \boldsymbol{F}[\Omega]\boldsymbol{\mathfrak{a}}^\text{in}[\Omega]
\end{equation}

We recall that we can go from the sideband basis to the quadrature basis using the isomorphism:
\begin{equation}
\boldsymbol{x}^\text{out} = \boldsymbol{G}\boldsymbol{x}^\text{in} = \boldsymbol{P F P^\dagger} \boldsymbol{x}^\text{in}
\end{equation}

where $\boldsymbol{P} = \frac{1}{\sqrt{2}} \begin{bmatrix}
1 & 1\\
-i & i
\end{bmatrix}$.

Finally, to make the spectral density matrix (SDM) compatible with the added quantum noise, the transformation is \cite{ding2024}(we use single sided power spectral densities)
\begin{equation}
\boldsymbol{\bar{S}}^\text{out} = \boldsymbol{G} \left(\boldsymbol{\bar{S}}^\text{in} - \boldsymbol{1}\right) \boldsymbol{G}^\dagger + \boldsymbol{1}
\end{equation}

Analytically, $\boldsymbol{G}[\Omega]$ has the expression
\begin{equation}
    \boldsymbol{G}[\Omega] = e^{\ii \theta_C} \left(C[\Omega] \boldsymbol{1} - D[\Omega] \boldsymbol{\sigma}\right) \boldsymbol{R}(\theta_D)
\end{equation}
\noindent where 
\begin{align}
    C[\Omega] &= \frac{|F[\Omega]| +|F[-\Omega]| }{2}\\
    D[\Omega] &=  \frac{|F[\Omega]| -|F[-\Omega]| }{2}\\
    \theta_C[\Omega] &= \frac{\arg(F[\Omega]) -\arg(F[-\Omega]) }{2}\\
    \theta_D[\Omega] &= \frac{\arg(F[\Omega]) + \arg(F[-\Omega]) }{2}\\
    \boldsymbol{\sigma} &= \begin{bmatrix}
        0 & -\ii \\
        \ii & 0
    \end{bmatrix}
\end{align}

We assume that the input SDM is a squeezed rotated state $ \boldsymbol{\bar{S}}^\text{in} = \boldsymbol{R}(\phi) \boldsymbol{S}(2r) \boldsymbol{R}(\phi)^T$,
 with $\boldsymbol{S}(r) = \text{diag}(e^{r}, e^{-r})$ the squeezing matrix. 
Noting that $\boldsymbol{R}(\theta_D)$ commutes with $\boldsymbol{\sigma}$ and that the latter matrix is hermitian, we find that

\begin{equation}
\begin{split}
\boldsymbol{\bar{S}}^\text{out} = \boldsymbol{R}(\theta_D + \phi) &\left[(C \boldsymbol{1} - D\boldsymbol{\sigma}) (\boldsymbol{S}(2r) - \boldsymbol{1}) (C \boldsymbol{1} - D\boldsymbol{\sigma}) \right.\\
&\quad \left.+\boldsymbol{1}\right]\boldsymbol{R}(\theta_D + \phi)^T
\end{split}
\end{equation}

The properties of this matrix are better understood by factoring out the rotation matrix, that is, define $\boldsymbol{\bar{S}'}\eqdef \boldsymbol{R}(\theta_D + \phi)^T \boldsymbol{\bar{S}}^\text{out} \boldsymbol{R}(\theta_D +\phi)$ so that 

\begin{equation}
\boldsymbol{\bar{S}'} = (C \boldsymbol{1} - D\boldsymbol{\sigma}) (\boldsymbol{S}(2r) - \boldsymbol{1}) (C \boldsymbol{1} - D\boldsymbol{\sigma}) +\boldsymbol{1}
\end{equation}

Explicitly, we find
\begin{widetext}
\begin{equation}\label{eq:SDM_prime}
    \boldsymbol{\bar{S}'} = \begin{bmatrix}
        \eta \left[(1-\Xi) e^{2r} + \Xi e^{-2r} \right] + (1-\eta)  & 4\ii CD \sinh^2(r)\\
        -4\ii CD \sinh^2(r) & \eta \left[\Xi e^{2r} + (1-\Xi) e^{-2r} \right] + (1-\eta) 
    \end{bmatrix}
\end{equation}

\end{widetext}

\noindent where 
\begin{align}
   \label{eq:eta}
   \eta &\eqdef C^2 + D^2 = \frac{|F[\Omega]|^2 + |F[-\Omega]|^2}{2}\\
    \label{eq:Xi}
    \Xi &\eqdef \frac{D^2}{\eta} = \frac{1}{2} - \frac{|F[\Omega]F[-\Omega]|}{|F[\Omega]|^2 + |F[-\Omega]|^2}
\end{align}

Thus the squeezing quadratures are rotated by an amount equal to the frequency-dependent angle $\theta_D[\Omega]$. Degradation mechanisms may affect this angle; we thus define misphasing $\Delta \theta_D[\Omega]$ as the difference between $\theta_D[\Omega]$ when there is no loss in the passive system and the lossy case:
\begin{equation}
    \Delta\theta_D[\Omega] \eqdef \theta_D[\Omega] \big|_\text{lossless} -\theta_D[\Omega]
\end{equation}

The expression \cref{eq:SDM_prime} helps one understand how squeezing is degraded when doing homodyne detection. The general expression of the homodyne photocurrent spectrum is \cite{ding2024}
\begin{equation}
    \bar{S}_{\theta}^\text{HD}[\Omega] = \boldsymbol{u}(\theta)^T \text{Re}(\boldsymbol{\bar{S}'}[\Omega])\boldsymbol{u}(\theta) = \Lambda_+^\mathbb{R}\cos^2(\theta) + \Lambda_-^\mathbb{R} \sin^2(\theta)
\end{equation}
\noindent where the homodyne readout is $\vect{u}(\theta) = \begin{bmatrix}
    \cos(\theta) \\
    \sin(\theta)
\end{bmatrix}$, and $\Lambda_\pm^\mathbb{R}\eqdef  \eta \left[(1-\Xi) e^{\pm 2r} + \Xi e^{\mp 2r} \right] + (1-\eta) $ are the eigenvalues of $\text{Re} (\boldsymbol{\bar{S}'})$. This can be seen by noting that it is a diagonal matrix per \cref{eq:SDM_prime}. 

The interferometer is designed to rotate the squeezed field by amount $\pi/2 - \theta_D[\Omega] \big|_\text{lossless}$. Effectively, this means that the equivalent homodyne angle is $\theta = \Delta\theta_D[\Omega] + \pi/2$, so the measured squeezing through homodyne detection is
\begin{equation}
    \bar{S}^\text{HD}[\Omega] = \Lambda_+^\mathbb{R} \sin^2(\Delta\theta_D[\Omega]) + \Lambda_-^\mathbb{R} \cos^2(\Delta\theta_D[\Omega])
\end{equation}
\noindent which is exactly \cref{eq:SQZ_measurement_general}.

\medskip

\subsection{Phase noise}\label{app:phase_noise}

Phase noise on a parameter $X_k$ on which the spectral density depends is accounted by effectively by averaging its expression in the interval $[X_k-\delta X_k, X_k+\delta X_k]$:
\begin{equation}
\begin{split}
\bar{S}^\text{avg} = &\bar{S}(X_1,\dots, X_n) \\
&+ \sum_{k} \left(\frac{\bar{S}(X_k + \delta X_k) + \bar{S}(X_k - \delta X_k)}{2} - \bar{S}(X_k )\right)\\
\end{split}
\end{equation}

\noindent where in the sum, the dependence of $\bar{S}$ on all the other parameters $X_j$ for $j\neq k$ are made implicit and evaluated at the setpoint \cite{Kwee_2014}.

\medskip
\section{Proof of the scaling laws for bandwidth, detuning and loss} \label{app:proof_scaling_laws}
In this appendix we prove the equations \cref{eq:gamma_c,eq:gamma_a,eq:Delta_c,eq:Delta_a} as well as the observation of \cref{sec:optical_loss}, that the responses of 2FC and CFC are similar when their RTL $\Lambda$ and $\Lambda'$ are equal. Our proof generalizes that of \cite{Zhang24} to this latter observation.

\subsection{Single filter cavity}
We recall that the frequency response of a single filter cavity
\begin{equation}
    r_\text{cav} = \frac{r- \sqrt{1-\Lambda }\, \ee^{-2\ii \varphi}}{1-{r}\sqrt{1-\Lambda}\, \ee^{-2\ii \varphi}}
\end{equation}
 where the input mirror amplitude reflectivity is $r$, the RTL is $\Lambda$, the single-trip phase is $\varphi \eqdef (\Delta \omega + \Omega) L_0/c = (\omega_0 \Delta L + \Omega L_0)/c $. Expanding at lowest order in the input mirror transmission $T$, $\Lambda$ and $\phi \eqdef -2\varphi$ gives $r =\sqrt{1-T} \simeq 1-\frac{T}{2}$, $\sqrt{1-\Lambda} \simeq 1-\frac{\Lambda}{2}$, $\ee^{-2\ii \varphi} \simeq 1+\ii \phi$. Hence
 \begin{equation}
     r_\text{cav} \simeq \frac{-T+\Lambda - 2\, \ii\phi}{ T+ \Lambda - 2\, \ii \phi}
 \end{equation}

\subsection{Two cavities}
The 2FC configuration is simply the product of the responses of two single cavities (indices 1 and 2). Assuming that the RTL $\Lambda\ll T_{1,2}$ correspond to mirror absorption loss, and that they are the same for all mirrors in the setup,
\begin{align}
    r_\text{2FC} &= \frac{-T_1+\Lambda - 2\, \ii\phi_1}{ T_1+ \Lambda - 2\, \ii \phi_1} \times \frac{-T_2+\Lambda - 2\, \ii\phi_2}{ T_2+ \Lambda - 2\, \ii \phi_2}\\
    \label{eq:r2FC_approx}
    &= \frac{\frac{T_1T_2}{4} + \frac{\ii}{2}[\phi_1(T_2-\Lambda) + \phi_2(T_1-\Lambda)]- \phi_1\phi_2}{\frac{T_1T_2}{4} -\frac{\ii}{2}[\phi_1(T_2+\Lambda)+ \phi_2(T_1+\Lambda)]- \phi_1\phi_2}
\end{align}

\subsection{Coupled Filter Cavity with symmetric loss}
For a coupled filter cavity, the response is
\begin{equation}
    r_\text{CFC} = \frac{r_c - r_\text{ref} \, \sqrt{1-\Lambda'} \, \ee^{-2\ii\varphi_c}}{1-r_\text{ref} \, r_c \sqrt{1-\Lambda'}\, \ee^{-2\ii\varphi_c}}
\end{equation}
\noindent 
where $r_\text{ref}$ is the response of the a-cavity, and we have written $\Lambda'$ the RTL. It is a priori different from $\Lambda$ but we assume it to be identical for both sub-cavities of the CFC system. Thus the reflectivity of the a-cavity is
\begin{equation}
    r_\text{ref} = \frac{r_a- \sqrt{1-\Lambda' }\, \ee^{-2\ii \varphi_a}}{1-{r_a}\sqrt{1-\Lambda'}\, \ee^{-2\ii \varphi_a}}
\end{equation}

We inject in the previous equation to obtain
\begin{widetext}
\begin{equation}
    r_\text{CFC} = \frac{r_c - r_a \sqrt{1-\Lambda'} \, (r_c \ee^{-2\ii\varphi_a} + \ee^{-2\ii\varphi_c}) + (1-\Lambda') \, \ee^{-2\ii\varphi_a}\, \ee^{-2\ii\varphi_c} }{1- r_a \sqrt{1-\Lambda'} \, ( \ee^{-2\ii\varphi_a} + r_c \, \ee^{-2\ii\varphi_c}) + (1-\Lambda') r_c \, \ee^{-2\ii\varphi_a}\, \ee^{-2\ii\varphi_c}}
\end{equation}

Using the lowest-order approximations, the numerator becomes, after simplification
\begin{equation}
    \text{Num}(r_\text{CFC}) = T_a + \frac{\ii}{2}[(T_c + T_a - \Lambda') \phi_a + (T_a - \Lambda')\phi_c] - \phi_a\phi_c
\end{equation}

Comparing this with the numerator of \cref{eq:r2FC_approx}, and identifying the real and imaginary parts, we get
\begin{align}
\label{eq:real_part}
    T_a - \phi_a\phi_c &= \frac{T_1T_2}{4} -{\phi_1\phi_2}\\
\label{eq:im_part}
    (T_c + T_a-\Lambda')\phi_a + (T_a-\Lambda') \phi_c &= \phi_1(T_2-\Lambda) + \phi_2(T_1-\Lambda)
\end{align}

We now expand the expressions of $\phi_i = -2(\Omega + \Delta \omega_i) \tau$ for $i\in\{1, 2, a, c\}$, while $\tau = L/c$ where $L$ is the length the cavity, assumed to be identical for all four cavities. We then identify both sides of \cref{eq:real_part,eq:im_part} by powers of $\Omega$ and assuming $T_a\ll T_c$:
\begin{align}
    \gamma_a  &= \tau_c(\gamma_1 \gamma_2 - \Delta\omega_1\Delta\omega_2 + \Delta\omega_a\Delta\omega_c)\\
    \Delta \omega_a + \Delta\omega_c &= \Delta\omega_1 +  \Delta\omega_2\\
    \label{eq:Lambda_diff_1}
    2(\gamma_c + \gamma_a - \gamma_2 - \gamma_1) &= \frac{\Lambda' - \Lambda }{\tau}\\
    \label{eq:Lambda_diff_2}
    4(\gamma_c \Delta\omega_a + \gamma_a \Delta\omega_c - \gamma_2\Delta\omega_1 - \gamma_1 \Delta\omega_2) &= \frac{(\Delta\omega_1 + \Delta\omega_2)(\Lambda'-\Lambda)}{\tau}
\end{align}
\end{widetext}

\noindent where the bandwidths $\gamma_i = \frac{cT_i}{4L}$ correspond to the ones in the lossless case. Thus, the scaling laws \cref{eq:gamma_c,eq:gamma_a,eq:Delta_c,eq:Delta_a} are obtained by solving the previous equations assuming $\Lambda' =0$ and $\Lambda = 0$. Then, we obtain in the lossy case that $\Lambda' = \Lambda$ by looking at \cref{eq:Lambda_diff_1,eq:Lambda_diff_2}, the left sides being zero. This concludes the proof of \cref{eq:gamma_c,eq:gamma_a,eq:Delta_c,eq:Delta_a} and formally justifies the observation of \cref{sec:optical_loss}.

\medskip
\section{Substrate loss of the middle mirror in CFC}\label{app:sub_loss}
\begin{figure}
    \centering
    \includegraphics[width=\linewidth]{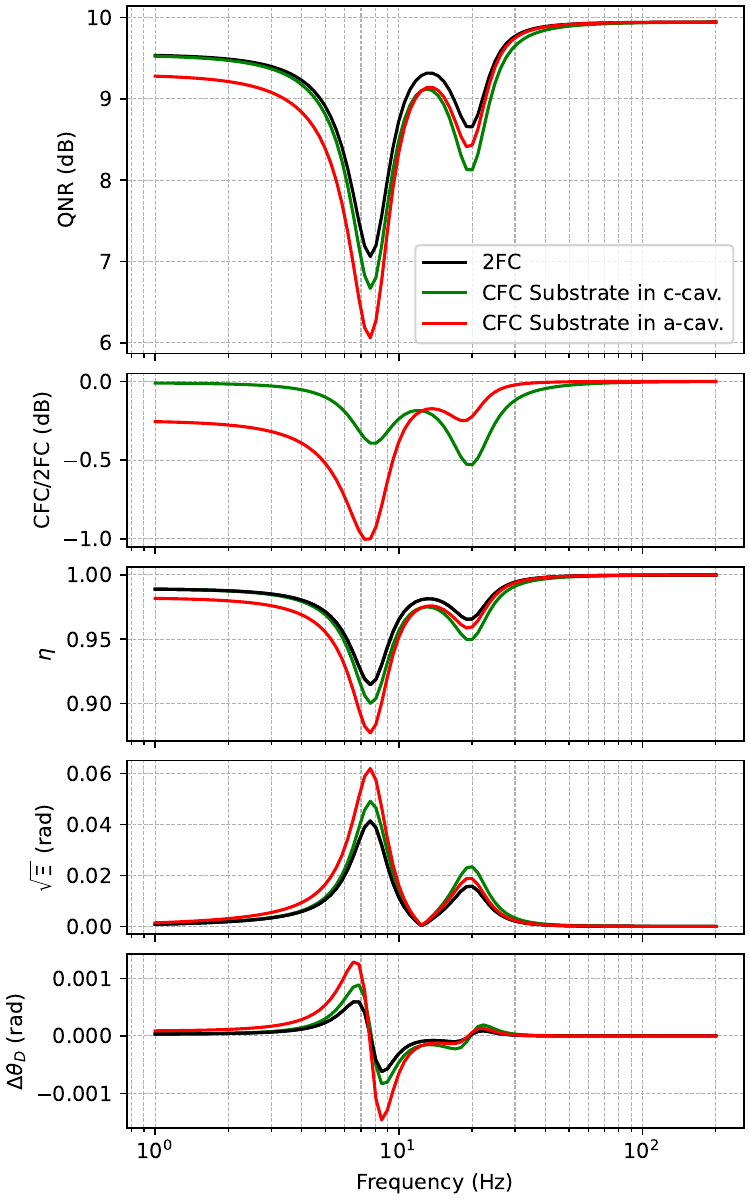}
    \caption{Comparison of CFC vs 2FC when only optical losses are considered, with extra 20 ppm round trip loss in CFC due to the substrate of the middle mirror.}
    \label{fig:sub_loss}
\end{figure}

In this appendix we justify the statement of \cref{sec:optical_loss} that substrate loss of the middle mirror in CFC has a negligible impact on the degradation of squeezing. We consider a middle mirror made of fused silica and 5 cm thick. Substrate loss in bulk fused silica depend on the dryness of the substrate and can be less than 1 ppm / cm \cite{Jerome,HUMBACH199619}. We retain a conservative 2 ppm/cm figure for bulk fused silica, which amounts to an added round trip loss of 20 ppm. Using this value, and adding the 30 ppm of loss per subcavity (see \cref{table:params_general}), we compare the squeezing performances when the substrate is facing towards the c-cavity (with High Reflectivity coating faced towards the a-cavity) to when it is facing the a-cavity. From the results in \cref{fig:sub_loss}, we see that by placing the substrate in the c-cavity (green curve), squeezing is never degraded by more than 0.5 dB compared to the 2FC scheme.

\medskip

\section{Geometrical defects in a coupled cavity}
\label{app:MMCFC}
In this section, we derive the mode mismatch amplitude for a coupled filter cavity with geometric imperfections, to justify that with current state-of-the-art optics, the intra-cavity mode mismatch $\Upsilon_a$ of \cref{table:params_general} can reasonably be neglected compared to the other mismatch contributions of this paper; in other words we show that $\Upsilon_a \ll 1\%$.

\begin{figure}
    \centering
    \includegraphics[width=1\linewidth]{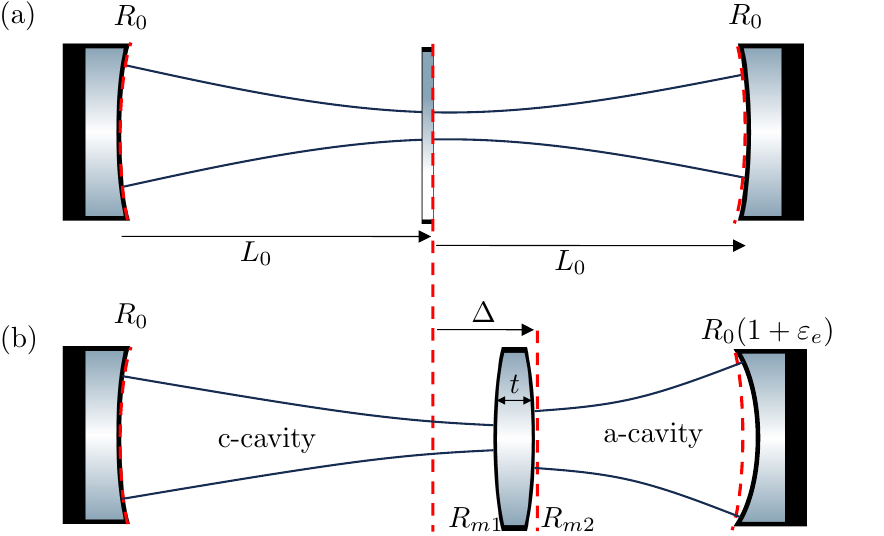}
    \caption{(a): Ideal coupled filter cavity: both input and end mirrors are concave with radius of curvature $R_0$; the middle mirror is plane and placed in the middle of the cavity, thus delimiting two sub-cavities of equal length. (b): Realistic coupled filter cavity, including radius error $\varepsilon_e$, middle mirror placement error $\Delta$, residual spheric interfaces $R_{m1}$, $R_{m2}$, and finite substrate thickness $t$ and index $n$ (the high-reflectivity surface is oriented towards the a-cavity).}
    \label{fig:imperfect_CFC}
\end{figure}

\subsection{Review of Gaussian optics}

 \subsubsection{Gaussian beams}
 The $q$ parameter is defined for a Gaussian beam propagating along $z$ as $q(z) = (z - z_0) + \ii z_R$ where $z_0$ is the position of the waist and $z_R$ is the Rayleigh range. Equivalently, we write
\begin{equation}
\frac{1}{q(z)} = \frac{1}{R(z)} - \frac{2\ii}{k w(z)^2}
\end{equation}
\noindent where $R(z) = \frac{(z-z_0)^2 + z_R^2}{z-z_0}$ is the spherical phase curvature of the beam and $w(z) = \sqrt{\frac{2}{kz_R}} \sqrt{(z-z_0)^2 + z_R^2}$ is the waist of the beam at position $z$. We define a normalized gaussian mode $u(r,z)$ of the electric field as
\begin{equation}
\begin{split}
    u(r,z) &= \sqrt{\frac{2}{\pi w(z)^2}} \exp\left[- \frac{\ii k r^2}{2 q(z)} \right] \exp\left[ \ii \varphi(z)\right] \\
    &= \sqrt{-\frac{k}{\pi} \Im\left(\frac{1}{q(z)}\right)} \exp\left[- \frac{\ii k r^2}{2 q(z)} \right] \exp\left[ \ii \varphi(z)\right] 
\end{split}
\end{equation}
\noindent where $\varphi(z) = -kz + \psi(z)$ with $\psi(z)$ the Gouy phase. This function obeys the normalization condition $\iint  |u(r,z)|^2 \dd x \dd y= 1$. This means that the amplitude of any electric field $E(r,z)$ with total transverse power $P_0$ can be written as $E(r,z) = \sqrt{P_0}\ u(r,z)$.

\subsubsection{Overlap integral and mode mismatch amplitude}

We are interested in the overlap integral between two modes $u_1$ and $u_2$ of parameters $q_1(z), q_2(z)$:
\begin{equation}
    \mathcal{O}(z) = \iint_{\mathbb{R}^2} u_1^*(r,z) u_2(r,z) \dd x \dd y
\end{equation}

\noindent This can be computed directly, by using the gaussian integral $\iint_{\mathbb{R}^2} e^{\ii a\frac{x^2 + y^2}{2}} \dd x \dd y = \frac{2\pi \ii}{a}$ if $\Im(a) >0$. Here we take $a = k\left(\frac{1}{q_1^*} - \frac{1}{q_2}\right)$ to get
\begin{equation}
    \mathcal{O}(z) = \frac{2\ii}{\frac{1}{q_1^*} - \frac{1}{q_2}} \sqrt{\Im\left(\frac{1}{q_1}\right)\Im\left(\frac{1}{q_2}\right)} \ e^{\ii (\varphi_2 - \varphi_1)}
\end{equation}
\noindent where we omitted the dependency in $z$ for clarity. Using the relation $\Im\left(\frac{1}{q}\right) = \frac{-z_R}{|q|^2}$, this simplifies further to
\begin{equation}
     \mathcal{O}(z) = 2\frac{\sqrt{z_{R1} z_{R2}}}{q_2 - q_1^*} \frac{\ii q_1^* q_2}{|q_1 q_2|} e^{\ii (\varphi_2 - \varphi_1)} 
\end{equation}
Note that the second fraction has modulus 1, so since we are only interested in the amplitude of this overlap integral, we get
\begin{equation}
    |\mathcal{O}(z)| = 2\frac{\sqrt{z_{R1} z_{R2}}}{|q_2 - q_1^*|}
\end{equation}
\noindent which is indeed independent of $z$ since $q_2 - q_1^* = z_{01} - z_{02} + \ii (z_{R2} + z_{R1})$. Finally, the modulus square is related to the mode matching as
\begin{equation}
    1-\Upsilon \eqdef |\mathcal{O}|^2 = \frac{4 z_{R1} z_{R2}}{|q_2 - q_1^*|^2}
\end{equation}
\noindent It is possible to obtain a compact expression of the mode mismatch amplitude $\Upsilon$ using the above expression at lowest order in $\Delta z_0 \eqdef z_{01} - z_{02}$ and $\Delta z_R \eqdef z_{R1} - z_{R2}$. Since $|q_2 - q_1^*|^2 \simeq \Delta z_0^2 + 4 z_R^2 \simeq 4 z_R^2$, we have
\begin{equation}\label{eq:mode_mismatch_rayleigh}
    \Upsilon \simeq \frac{\Delta z_0^2}{4 z_R^2} + \frac{\Delta z_R^2}{4 z_R^2}
\end{equation}
Equivalently, in terms of waists, we use $z_{Ri} = k w_{0i}^2/2$ to obtain $\Upsilon = {\frac{\Delta z_0^2}{4 z_R^2} + \frac{\Delta w_0^2}{w_0^2}}$, where $\Delta w_0 = w_{01} - w_{02}$.

\subsubsection{Fundamental gaussian mode in a linear cavity}
In the rest of this appendix, the longitudinal coordinate grows left to right. 
For a two-mirror cavity of length \(L\) with input/end mirror radii of curvature \(R_i,R_e\) (using the convention that, seen from inside the cavity, a mirror has positive radius of curvature if it is concave), the waist position $z_0$ \emph{measured from the first mirror} (\(z=0\) at the mirror $R_i$), and its Rayleigh length squared $z_R^2$ are
	\begin{align}
		z_0 &= \frac{L^2 - R_e L}{\,2L - R_i - R_e\,}, \label{eq:z0-gen}\\[2pt]
		z_R^2 &= \frac{L (R_i - L)(R_e - L) (R_i + R_e - L)}{(R_e + R_i - 2L)^2}. \label{eq:zR2-gen}
	\end{align}

Note that for a perfect, flat–curved half-cavity (first mirror plane, second mirror \(R_0\), length \(L_0\)), \cref{eq:z0-gen,eq:zR2-gen} give $z_0^{(0)}=0$ and $z_R^2=L_0(R_0-L_0)$.

\subsection{Mode mismatch due to geometrical imperfections}
Consider first the ideal situation shown in \cref{fig:imperfect_CFC} (a), of a symmetric Fabry--Perot cavity of length \(2L_0\) with end mirrors of the same radius \(R_0\) and a plane mirror in the middle.  
	Each sub-cavity (``c-cavity'' and ``a-cavity'') is stable with perfect fundamental Gaussian having its waist at the plane mirror and
	\begin{equation}
		z_R=\sqrt{L_0\,(R_0-L_0)}. 
	\end{equation}
	
	We now add realistic imperfections, as shown in \cref{fig:imperfect_CFC} (b):
	\begin{itemize}
		\item[(i)] Right end-mirror curvature deviation: \(R_e=R_0(1+\varepsilon_e)\), with \(|\varepsilon_e|\ll 1\);
		\item[(ii)] Middle plane displaced: left gap \(L_1=L_0+\Delta\), right gap \(L_2=L_0-\Delta\), with \(|\Delta|\ll L_0, R_0\);
		\item[(iii)] Middle optic: entrance surface curvature \(R_{m1}\), back (HR-coated) surface curvature \(R_{m2}\), both large in magnitude compared to all other radii of curvature;
		\item[(iv)] Middle substrate: thickness \(t\ll \Delta, L_0, R_0\), refractive index \(n>1\); the HR coating is on the back surface, so the left half-beam traverses the glass to reflect.
	\end{itemize}

    Note that since we are only interested in the relative mode mismatch between the two eigenmodes of the two sub-cavities, we have assumed without loss of generality that the input mirror is undeformed. Additionally, the impact of coating non-uniformity can be neglected compared to these four imperfections, as the coatings are much thinner than any other length scale above.
	
	Label by $c$ (left) and $a$ (right) the fundamental Gaussian eigenmodes of the two imperfect half-cavities.  
	We compute, to the lowest non-trivial order in the small quantities \(\varepsilon_e,\Delta,\,1/R_{m1},\,1/R_{m2},\,t\), the mode mismatch between the two eigenmodes. 
	We will repeatedly use the geometry factors
	\begin{equation}
    \begin{split}
		c_\Delta &:= \frac{L_0-\tfrac{R_0}{2}}{z_R},\\
		c_k &:= \frac{z_R}{2}\,(2L_0-R_0),\\
		c_\varepsilon &:= \frac{R_0\,z_R}{2\,(R_0-L_0)}.
		\label{eq:geom-coefs}
\end{split}\end{equation}

	\subsubsection{a-cavity}
	This cavity is bounded by the middle HR surface (first mirror) and the right end mirror, so
	$L=L_0-\Delta$, $R_i=R_{m2}$ (large) and $R_e=R_0(1+\varepsilon_e)$.

	Evaluating \eqref{eq:z0-gen} with large \(R_i\):
	\begin{equation}
	z_{0a}
	= -\,\frac{L^2-R_e L}{R_i} = \frac{z_R^2}{R_{m2}}
	\end{equation}
    \noindent where we neglected terms of order $R_i^{-2}$. This is the position of the waist relative to the position of the HR surface of the middle mirror, itself displaced by $\Delta$ from the midpoint of the CFC. Thus the waist position relative to the midpoint of the CFC is 
    \begin{equation}\label{eq:z_02}
    \tilde{z}_{0a} = \,z_{0a} + \Delta = \Delta + \frac{z_R^2}{R_{m2}}
    \end{equation}
	
    Similarly, \eqref{eq:zR2-gen} reduces to lowest order in $R_i^{-1}$ to
	\begin{align}
	z_{Ra}^2 &= L (R_e - L) \;-\; \frac{L (R_e - L)(R_e-2L)}{R_i}\\
    & = z_{R}^2 - (R_0 - 2L_0) \Delta  + L_0 R_0 \varepsilon_e + z_R^2 \frac{2L_0-R_0}{R_{m2}}
	\end{align}
	This means that the Rayleigh length is
	\begin{equation}
		z_{Ra} = z_R +  c_\Delta\,\Delta + c_\varepsilon\,\varepsilon_e + \frac{c_k}{R_{m2}}.
		\label{eq:z_R2}
	\end{equation}

    	\subsubsection{c-cavity}
	
	The left cavity is bounded by the \emph{input} mirror (first mirror \(R_0\)) and the \emph{effective} reflector seen in vacuum when the beam enters the substrate, reflects from the HR on the back surface, and exits. The effect of the substrate is to add optical path and change the radius of curvature of the high-reflective surface, such that, to first order, $L=L_0+\Delta + \frac{t}{n} = L_0 + \tilde{\Delta}$, $R_i=R_0$ and $R_e=\frac{-R_{m2}}{n}$ (the minus sign is because $+R_{m2}$ corresponds to the radius of curvature as seen from the a-cavity, outside the c-cavity). Note that $R_{m1}$ has no contribution at lowest order because the beam crosses from vacuum to substrate then back.

	Expanding \eqref{eq:z0-gen} for large \(|R_e|\):
	\begin{equation}
	z_{0c} 
	= L + \frac{L(L-R_i)}{R_e} = L_0 + \tilde{\Delta} + \frac{nz_R^2}{R_{m2}}.
	\end{equation}
This is the position of the waist relative to the input mirror, which is $L_0$ to the left of the midpoint of the CFC. This means that, relative to the midpoint, the waist is at position 
\begin{equation}\label{eq:z_01}
\tilde{z}_{0c} = z_{0c} -L_0 = \Delta + \frac{t}{n} + \frac{nz_R^2}{R_{m2}}
\end{equation}

For the Rayleigh range, we use \eqref{eq:zR2-gen} with large \(|R_e|\) to get:
	\begin{equation}
	z_{Rc}^2 
	= L (R_i - L) \;-\; \frac{L (R_i - L)(R_i-2L)}{R_e}.
	\end{equation}
With $L = L_0 + \tilde{\Delta}$ and $R_i = R_0$, we obtain
\begin{equation}\label{eq:z_R1}
    z_{Rc} = z_R - c_\Delta \left(\Delta + \frac{t}{n}\right) - \frac{nc_k}{R_{m2}}
\end{equation}

\subsubsection{Application}

We apply the previous formulas to our coupled filter cavity system, in which $R_0 \sim 15$ km, $L_0 \sim 5$ km. We choose a residual radius of curvature $R_{m2} \sim 10^6$ m and the substrate has index $n \sim 1.4$ with thickness $t \sim 5$ cm and position error $\Delta \sim 1$ m. This means that the reference Rayleigh range is $z_R = \sqrt{L_0(R_0-L_0)} = 7.07$ km and the relative waist position between the two modes is $\abs{\frac{\Delta z_0}{2z_R}} = \abs{\frac{t}{2nz_R} - \frac{(n-1) z_R}{2R_{m2}}} \sim 1\cdot 10^{-3}$  (using \cref{eq:z_01,eq:z_02}) while the relative Rayleigh range between the two modes is 
$\abs{\frac{\Delta z_R}{2z_R}} = \abs{c_k\frac{-n-1}{2 z_RR_{m2}} -\frac{c_\Delta}{2 z_R} \left(2\Delta + \frac{t}{n}\right) - \frac{c_\epsilon}{2z_R} \epsilon_e} \sim 
3\cdot 10^{-3}$ (using \cref{eq:z_R1,eq:z_R2}), which leads to a mode mismatch (see \cref{eq:mode_mismatch_rayleigh}) $\Upsilon_a \sim 1 \cdot10^{-5}$ that is three orders of magnitude smaller than any other mode matching degradation considered in the main text. In conclusion, this justifies the claim that $\Upsilon_a \sim 0$ in \cref{table:params_general}.

\section{Addition of mode mismatches}
\label{app:add_mm}
Let $\Upsilon_A, \psi_A$ and $\Upsilon_B, \psi_B$ be mismatch parameters. Let $\boldsymbol{U}_C \eqdef \boldsymbol{U}(\Upsilon_A, \psi_A)\boldsymbol{U}(\Upsilon_B, \psi_B)$. We wish to write $\boldsymbol{U}_C$ in a form similar to $\boldsymbol{U}(\Upsilon', \psi')$ where $(\Upsilon', \psi')$ can be expressed as a function of $\Upsilon_A, \psi_A, \Upsilon_B, \psi_B$.

The product yields
    \begin{equation}
        \boldsymbol{U}_C = \begin{bmatrix}
            \sqrt{1- \Upsilon'}\, e^{\ii\psi_0} & - \sqrt{\Upsilon'}\,  e^{\ii \psi_1} \\
            \sqrt{\Upsilon'} \, e^{-\ii \psi_1} & \sqrt{1- \Upsilon'} \, e^{-\ii\psi_0} 
        \end{bmatrix}
    \end{equation}
\noindent where 
\begin{equation}\label{eq:Upsilon_prime}
\begin{split}
    \Upsilon' = &\Upsilon_A + \Upsilon_B - 2\Upsilon_A \Upsilon_B \\
    &+ 2 \sqrt{\Upsilon_A(1-\Upsilon_B) \Upsilon_B(1-\Upsilon_A)} \, \cos(\psi_A - \psi_B)
    \end{split}
\end{equation}
\noindent $\psi_0 = \arg(\sqrt{(1-\Upsilon_A)(1-\Upsilon_B)} - \sqrt{\Upsilon_A\Upsilon_B}\,  e^{\ii (\psi_A -\psi_B)})$ and $\psi_1 = \arg(\sqrt{(1-\Upsilon_A) \Upsilon_B} \, e^{\ii \psi_B} + \sqrt{(1-\Upsilon_B) \Upsilon_A} \, e^{\ii \psi_A})$. Defining $\psi' \eqdef \psi_1 - \psi_0$, we get that 
\begin{equation}
    \boldsymbol{U}_C = e^{\ii \psi_0} \, \boldsymbol{U}(\Upsilon', \psi') \, \text{diag}(1, e^{-2\ii\psi_0})
\end{equation}

Note that on the right hand side, the complex exponential adds a global phase so is irrelevant, while the diagonal matrix's non trivial bottom right component is usually applied to the higher order mode in a vacuum state, whose properties do not depend on the phase either, so can also be neglected. Under these assumptions, we thus have $\boldsymbol{U}_C \sim \boldsymbol{U}(\Upsilon', \psi')$.

We see that $\Upsilon'$ is maximized when $\psi_A = \psi_B$ (constructive interference of the higher order mode) and equals
\begin{equation}
    \Upsilon'_\text{max} = \left(\sqrt{\Upsilon_A(1-\Upsilon_B)} + \sqrt{\Upsilon_B(1-\Upsilon_A)}\right)^2
\end{equation}
\noindent while its minimum is attained for $\psi_A = \psi_B + \pi$ and equals
\begin{equation}
     \Upsilon'_\text{min} = \left(\sqrt{\Upsilon_A(1-\Upsilon_B)} - \sqrt{\Upsilon_B(1-\Upsilon_A)}\right)^2   
\end{equation}

These expressions can be simplified by defining the polar variables $\theta_A$ and $\theta_B$ such that $\sin(\theta_A) = \sqrt{\Upsilon_A}$, $ \cos(\theta_A) = \sqrt{1-\Upsilon_A}$ and similarly for $\theta_B$. Then, 
\begin{equation}
    \Upsilon'_\text{max} = \sin^2(\theta_A + \theta_B)  \quad \text{and}\quad \Upsilon'_\text{max} = \sin^2(\theta_A - \theta_B)
\end{equation}

\medskip
\section{Quantum transfer function of ET-LF}
\label{app:ITF_response}

In this section we derive the quantum response of ET-LF including losses at the input, in the signal-recycling cavity, in the arms, and at the readout. We first recall the elementary components of the formalism, then we build up the full setup of ET-LF. 

\subsection{Matrix formalism for active systems}
Because ET-LF has a pondermotive response due to the radiation pressure acting on the cavity's mirrors, we will need to consider both sidebands when computing its input-output relation. As much as possible, we will work with the quadrature vectors $\boldsymbol{x}[\Omega] \eqdef [\hat{q}[\Omega], \ \hat{p}[\Omega]]^T$. For simplicity, we only consider a single mode here.

\subsubsection{Propagation of length $L$}
We represent a propagation over a length $L$ by the matrix $\Propag_L$ such that $\boldsymbol{x}^\text{out} = \Propag_L \boldsymbol{x}^\text{in}$ with
\begin{equation}\label{eq:propag}
\Propag_L = e^{\ii \Omega \tau} \Rot(\phi)
\end{equation}
\noindent with $\tau = L/c$ and $\phi = \omega_0\tau$. 

\subsubsection{Radiation pressure}
Consider a laser field of power $P$ impinging on a free-mass mirror of unit reflectivity, of mass $m$. The gravitational wave acts on this mirror with strain $h$ relative to a length $L$. Then
\begin{equation}\label{eq:TF_RP}
    \boldsymbol{x}^\text{out} = \RP\, \Xin + \boldsymbol{v}\,  h
\end{equation}

\noindent where 
\begin{equation}
    \RP = \begin{bmatrix}
        1 & 0\\
        -K & 1
    \end{bmatrix}, \quad \boldsymbol{v} = \frac{\sqrt{2K}}{h_\text{SQL}} \begin{bmatrix}
        0 \\
        1
    \end{bmatrix}
\end{equation}
\noindent with the Kimble factor $K = \frac{8 P \omega_0}{mc^2 \Omega^2}$ and $h_\text{SQL} = \sqrt{\frac{8\hbar}{m \Omega^2 L^2}}$. If there are resonance conditions on the impinging power  --- say we are considering intra-cavity power, then $P$ may also depend on frequency in general. Also note that if the reflection has the minus sign convention, then we need to replace $\RP$ by $-\RP$. 

\subsubsection{Retroaction on fixed mirror}
\begin{figure}
    \centering
    \includegraphics[width=\linewidth]{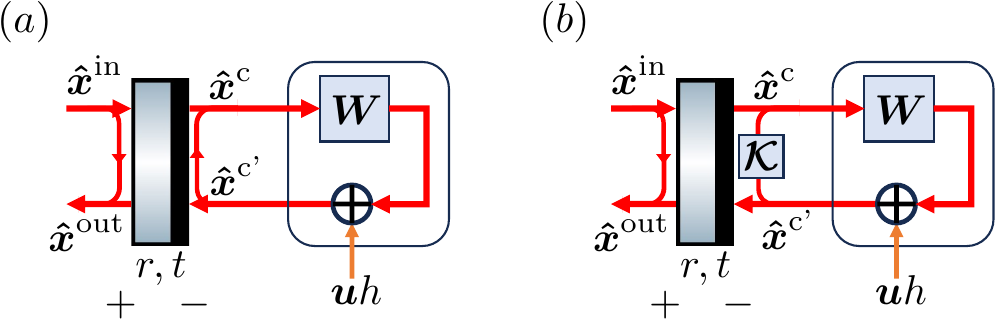}
    \caption{Block diagram of retroaction on a fixed (a) or mobile (b) mirror. The frequency-dependent matrix $\boldsymbol{W}$ represents the internal transfer function while the frequency-dependent vector $\boldsymbol{u}$ represents the coupling to an external signal $h$.}
    \label{fig:retroaction_diagram}
\end{figure}
Consider a cavity-type system, composed of an input mirror $(r,t)$ (minus side oriented inside of the cavity) and an internal transfer function matrix $\Wmat$ with possible coupling to the strain through a term in $\vect{u} h$, as depicted in \cref{fig:retroaction_diagram} (a). This means that the intra-cavity and output fields have the relations $\boldsymbol{x}^\text{c'} = \Wmat \boldsymbol{x}^\text{c} + \vect{u} h, 
    \boldsymbol{x}^\text{c} = t \Xin -r \boldsymbol{x}^\text{c'},
    \Xout =  r \Xin + t \boldsymbol{x}^\text{c'}$. From this we deduce the transfer function
\begin{equation}\label{eq:TF_retro}
    \Xout = (\boldsymbol{1} +r \Wmat )^{-1} (r+ \Wmat) \Xin + t(\boldsymbol{1} + r\Wmat)^{-1} \vect{u}h
\end{equation}

Notice that if the input mirror's minus side were oriented towards the outside of the cavity, then the input-output relation would be identical to the one above under the swap $r \to -r$.

\subsubsection{Retroaction on mobile mirror}
For the ET arm cavities, both the input and output mirrors are free masses, so they are both affected by radiation pressure noise. In this case, and neglecting the radiation pressure coming from the input field on the cavity (because the finesse is high), one obtains the set of equations $
    \boldsymbol{x}^\text{c'} = \Wmat \boldsymbol{x}^\text{c} + \vect{u} h$, $ 
    \boldsymbol{x}^\text{c} = t \Xin -r R \RP \boldsymbol{x}^\text{c'}$, 
    $\Xout =  r \Xin + t \boldsymbol{x}^\text{c'}$, because a fraction $R= r^2$ of the intra-cavity light pushes the input mirror towards the incident beam (ponderomotive $\RP$). Taking the limit $R \to 1$ (but keeping $r$ and $t$ as they are) and inverting, one obtains
\begin{equation}\label{eq:TF_retro_RP}
\begin{split}
    \Xout = &[r + (t^2 + r^2\RP) \Wmat](\boldsymbol{1} + r \RP \Wmat)^{-1} \Xin \\
    &+ t[ \boldsymbol{1} - r \Wmat (1+r \RP \Wmat)^{-1} \RP] \vect{u}h
\end{split}
\end{equation}

This situation is illustrated in \cref{fig:retroaction_diagram} (b).

\subsubsection{Retroaction with intra-cavity loss}
\begin{figure}[h]
    \centering
    \includegraphics[width=0.7\linewidth]{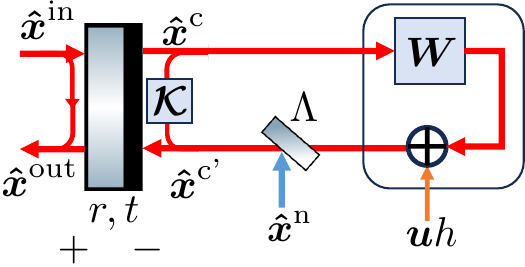}
    \caption{Retroaction with round-trip loss $\Lambda$, coupling vacuum noise $\boldsymbol{x}^\text{in}$.}
    \label{fig:retroaction_loss}
\end{figure}

Now, consider a cavity-type system with round-trip losses $\Lambda$, as shown in \cref{fig:retroaction_loss}. They are modeled by a beam-splitter right before the return intra-cavity field $\boldsymbol{x}^\text{c'}$, which leaks some vacuum noise $\boldsymbol{x}^\text{n}$ into the system. That is, if the (lossless) internal dynamics of the intra-cavity field is $\Wmat$ and the coupling to the GW strain is $\boldsymbol{u}$, we have the modified internal relation
\begin{equation}
    \vect{x}^\text{c'} = \sqrt{1-\Lambda}(\Wmat \boldsymbol{x}^\text{c} + \boldsymbol{u} h) + \sqrt{\Lambda} \boldsymbol{x}^\text{n}
\end{equation}
This is formally equivalent to the lossless relation
\begin{equation}
    \vect{x}^\text{c'} = \Wmat'\boldsymbol{x}^\text{c} + \boldsymbol{u}' h
\end{equation}
\noindent under the redefinition
\begin{equation}
    \Wmat' = \sqrt{1-\Lambda}\Wmat \quad\text{and}\quad \boldsymbol{u'} = \sqrt{1-\Lambda}\,  \boldsymbol{u} + \sqrt{\Lambda} \frac{\boldsymbol{x}^\text{n}}{h} 
\end{equation}

Thus, explicitly, the input-output relation on a fixed mirror (minus side inside) is
\begin{equation}\label{eq:TF_retro_lossy}
\begin{split}
    \Xout = &(\boldsymbol{1} +r \sqrt{1-\Lambda}\Wmat )^{-1} (r+ \sqrt{1-\Lambda}\Wmat) \Xin \\
    &+ t\sqrt{1-\Lambda}(\boldsymbol{1} + r\sqrt{1-\Lambda}\Wmat)^{-1} \vect{u}h\\
    & + t\sqrt{\Lambda}(\boldsymbol{1} + r\sqrt{1-\Lambda}\Wmat)^{-1} \vect{x}^\text{n}
\end{split}
\end{equation}

\noindent while for a free-mass input mirror it is
\begin{equation}
    \label{eq:TF_retro_free_lossy}
    \begin{split}
        \Xout &= (\boldsymbol{1} +r \sqrt{1-\Lambda}\RP\Wmat )^{-1} (r+ \sqrt{1-\Lambda}\RP\Wmat) \Xin \\
        & + t \sqrt{1-\Lambda} [ \boldsymbol{1} - r \sqrt{1-\Lambda}\, \Wmat (\boldsymbol{1}+r \sqrt{1-\Lambda}\RP \Wmat)^{-1} \RP] \vect{u}h\\
        & + t \sqrt{\Lambda} [ \boldsymbol{1} - r \sqrt{1-\Lambda}\, \Wmat (\boldsymbol{1}+r \sqrt{1-\Lambda}\RP \Wmat)^{-1} \RP] \vect{x}^\text{n} 
    \end{split}
\end{equation}

\subsection{Application to a Fabry-Pérot cavity}
\label{sec:lossy_FP_response}

\begin{figure}[h]
    \centering
    \includegraphics[width=0.7\linewidth]{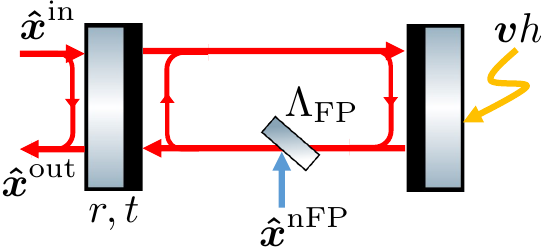}
    \caption{Physical diagram of a Fabry-Pérot interferometer coupled to some GW signal $h$. Round trip losses $\Lambda_\text{FP}$ couple the intra-cavity fields to vacuum noise $\boldsymbol{x}^\text{nFP}$.}
    \label{fig:FP_diagram}
\end{figure}

We apply the previous results to a tuned Fabry-Pérot cavity interferometer illustrated in \cref{fig:FP_diagram}, with round trip loss $\Lambda_\text{FP}$. We allow the input mirror to freely move as it is the case for the arms in the ET-LF interferometer. We derive the input-output relation.

The internal transfer function is $\Wmat = -\Propag \RP \Propag$ and coupling is $\vect{u} = \Propag \vect{v}$ where the minus sign reflects the orientation of the coating of the end mirror and $\Propag = e^{\ii\Omega \tau}$ is the single-trip propagation matrix defined in \cref{eq:propag} for a tuned cavity. Using \cref{eq:TF_retro_free_lossy} yields
\begin{equation}\label{eq:TF_FP_lossy}
    \Xout = \RP_\text{FPlossy} \Xin + \boldsymbol{v}_\text{FPlossy} h + \boldsymbol{\mathcal{L}}_\text{FP} \boldsymbol{x}^\text{nFP}
\end{equation}
\noindent where 
\begin{align}
    \RP_\text{FPlossy} &=  (\boldsymbol{1} -r \sqrt{1-\Lambda_\text{FP}} e^{2\ii\phi} \RP^2 )^{-1} \nonumber\\
    & \qquad \times (r- \sqrt{1-\Lambda_\text{FP}}e^{2\ii\phi}\RP^2)\\
    \boldsymbol{v}_\text{FPlossy} &=  t \sqrt{1-\Lambda_\text{FP}} (\boldsymbol{1}-r \sqrt{1-\Lambda_\text{FP}}e^{2\ii\phi}\RP^2 )^{-1} \vect{u}\\
    \boldsymbol{\mathcal{L}}_\text{FP} &= t \sqrt{\Lambda_\text{FP}} (\boldsymbol{1}-r \sqrt{1-\Lambda_\text{FP}}e^{2\ii\phi}\RP^2 )^{-1}
\end{align}

We recall that the Kimble factor intervening in $\RP$ here is the one of pure radiation pressure 
$K =\frac{8 P_\text{cav} \omega_0}{mc^2\Omega^2}$ but with $P_\text{cav}$ the classical power circulating in the arms (after accounting for losses).

\subsection{Application to Fabry-Pérot Michelson interferometers}
\begin{figure}
    \centering
    \includegraphics[width=1\linewidth]{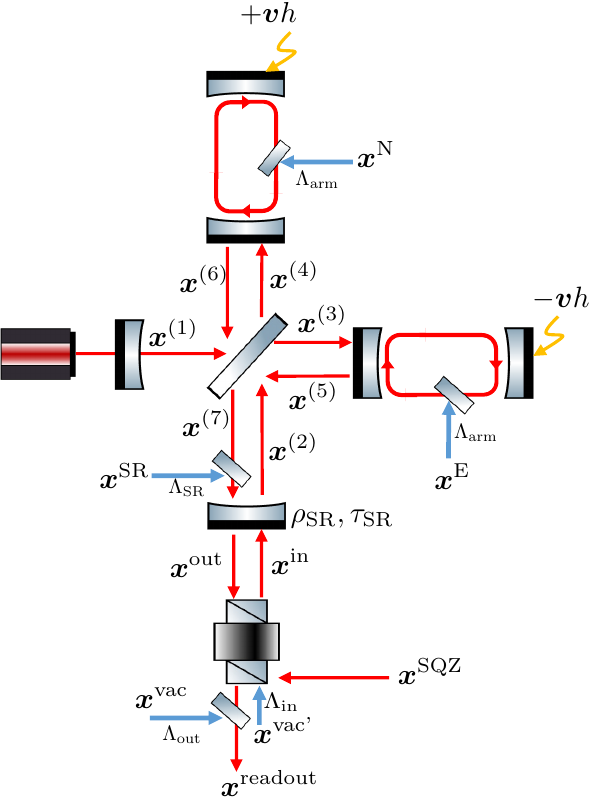}
    \caption{Physical diagram of a Dual-recycled, Fabry-Pérot Michelson Interferometer. Frequency-dependent squeezing can be injected through the field $\boldsymbol{x}^\text{SQZ}$ and the detected optical field is $\boldsymbol{x}^\text{readout}$. Losses enter the arms $\Lambda_\text{arm}$, the signal recycling cavity $\Lambda_\text{SR}$, the injection of squeezed vacuum $\Lambda_\text{in}$ and the readout $\Lambda_\text{out}$. }
    \label{fig:IFO_full_diagram}
\end{figure}

In this section we progressively build the full response of a lossy, dual-recycled Fabry-Pérot Michelson interferomter, depicted in \cref{fig:IFO_full_diagram}, starting from the response of a lossy interferometer without the signal recycling mirror, then nesting it with the signal recycling mirror and adding the corresponding losses. The configuration is shown in \cref{fig:IFO_full_diagram}, where for simplicity the arms are at a relative 90$^o$ angle, such that the GW strain $h$ is directly equal to the "differential arm length" degree of freedom. Note that for a 60$^o$ angle (as would be each interferometer in the triangular ET-LF configuration), the calculations are identical provided that the GW strain gets scaled by a factor of $\sin(60^o) = \sqrt{3}/2$ (\cite{Regimbau_2012}).

\subsubsection{Response of a lossy Power-recycled Fabry-Perot Michelson interferometer}
First notice that because the interferometer is tuned on dark fringe, power-recycling only affects the classical circulating power in the arms. Thus the response of the full system reduces to the response of a Simple Michelson with arm mirrors replaced by the FP cavity response of \cref{eq:TF_FP_lossy}, relating the north input $(4)$ to the output $(6)$ and the east input $(3)$ to its output $(5)$. Numbering the other fields as: the bright port $(1)$, the input dark port $(2)$, and the dark port output $(7)$; we obtain
\begin{align}
    \vect{x}^{(3)} &= \frac{\vect{x}^{(1)} - \vect{x}^{(2)}}{\sqrt{2}}\\
    \vect{x}^{(4)} &= \frac{\vect{x}^{(1)} + \vect{x}^{(2)}}{\sqrt{2}}\\
    \vect{x}^{(7)} &= \frac{\vect{x}^{(6)} - \vect{x}^{(5)}}{\sqrt{2}}
\end{align}

The North and East cavity arms have identical round-trip loss $\Lambda_\text{N} = \Lambda_\text{E} = \Lambda_\text{arm}$ (corresponding vacuua $\boldsymbol{x}^\text{N}$ and $\boldsymbol{x}^\text{E}$). Solving these equations, we obtain the interferometer's intput-output relation, with $\Xout = \vect{x}^{(7)}$ and $\Xin = \vect{x}^{(2)}$
\begin{equation}
    \Xout = \RP_\text{arm} \Xin - \sqrt{2}\, \vect{v}_\text{arm}h +  \boldsymbol{\mathcal{L}}_\text{arm} \boldsymbol{x}^\text{arm}
\end{equation}

\noindent where $\RP_\text{arm}, \vect{v}_\text{arm}, \boldsymbol{\mathcal{L}}_\text{arm}$ refer to $\RP_\text{FPlossy}, \boldsymbol{v}_\text{FPlossy}, \boldsymbol{\mathcal{L}}_\text{FP}$ of \cref{sec:lossy_FP_response}, and the vacuum quadrature vector $\boldsymbol{x}^\text{arm} = \frac{\boldsymbol{x}^\text{N} + \boldsymbol{x}^\text{E}}{\sqrt{2}}$. In particular, one can show that the explicit transfer function, at lowest order in the loss, corresponds exactly to equations (97)--(101) of \cite{klmtv}.

We can rewrite this transfer function as
\begin{equation}
    \Xout = \RP_\text{arm} \Xin + \boldsymbol{u}_\text{IFOlossy} h
\end{equation}
\noindent where 
\begin{equation}
 \boldsymbol{u}_\text{IFOlossy} = -\sqrt{2}\boldsymbol{v}_\text{arm} +  \boldsymbol{\mathcal{L}}_\text{arm} \frac{\boldsymbol{x}^\text{arm}}{h}
\end{equation}

\subsubsection{Response of a lossy, dual-recycled Fabry-Perot Michelson interferometer}
We nest the previous transfer function with the signal extraction mirror. We also add some internal loss $\Lambda\SR$ (noise $\boldsymbol{x}^\text{SR}$). We use  \cref{eq:TF_retro_lossy} (with the other mirror convention) with $\Wmat =  \Propag_\text{SR}\RP_\text{arm}\Propag_\text{SR}$, $\boldsymbol{u} = \Propag_\text{SR}\boldsymbol{u}_\text{IFOlossy}$. Finally, we can further split the input losses as $\Xin = \sqrt{1-\Lambda_\text{in}} \boldsymbol{x}^\text{SQZ} + \sqrt{\Lambda_\text{in}} x^\text{vac'}$ where only the term $\boldsymbol{x}^\text{SQZ}$ corresponds to a squeezed field, and we can incorporate some readout losses as $\boldsymbol{x}^\text{readout} = \sqrt{1-\Lambda_\text{out}} \boldsymbol{x}^\text{out} + \sqrt{\Lambda_\text{out}} x^\text{vac}$. Incorporating the input loss in $\vect{x}^\text{in}$ yields
\begin{equation}
\begin{split}
    \boldsymbol{x}^\text{readout} &= \boldsymbol{v}^h h + \boldsymbol{T}^\text{in} \Xin + \boldsymbol{T}^\text{arm} \vect{x}^\text{arm} \\
    & \quad + \boldsymbol{T}^\text{SR} \vect{x}^\text{SR} + \boldsymbol{T}^\text{vac} x^\text{vac}
    \end{split}
\end{equation}

\noindent where
\begin{align}
    \boldsymbol{v}^h &= -\sqrt{1-\Lambda_\text{out}}\sqrt{2}\, \tau\SR\sqrt{1-\Lambda\SR}\boldsymbol{A}\SR^{-1} \Propag_\text{SR}\vect{v}^\text{arm}\\
    \boldsymbol{T}^\text{in} &= \sqrt{1-\Lambda_\text{out}} \boldsymbol{A}\SR^{-1} \nonumber \\
    & \qquad \times (\rho\SR+ \sqrt{1-\Lambda\SR}\Propag_\text{SR}\RP_\text{arm}\Propag_\text{SR})\\
    \boldsymbol{T}^\text{arm}&= \sqrt{1-\Lambda_\text{out}}\tau\SR\sqrt{1-\Lambda\SR}\boldsymbol{A}\SR^{-1}\Propag_\text{SR}\boldsymbol{\mathcal{L}}_\text{arm}\\
    \boldsymbol{T}^\text{SR} &= \sqrt{1-\Lambda_\text{out}}\tau\SR\sqrt{\Lambda\SR}\boldsymbol{A}\SR^{-1}\\
    \boldsymbol{T}^\text{vac} &= \sqrt{\Lambda_\text{out}} \boldsymbol{1}
    \end{align}
\begin{align}
    \boldsymbol{A}\SR &= \boldsymbol{1} + \rho\SR\sqrt{1-\Lambda\SR}\Propag_\text{SR}\RP_\text{arm}\Propag_\text{SR}\\
    \Propag\SR &= e^{\ii\Phi\SR}\Rot(\phi_\text{SEC})
\end{align}
\begin{align}
    \RP_\text{arm} & = \boldsymbol{A}_\text{arm}^{-1} (r- \sqrt{1-\Lambda_\text{arm}}e^{2\ii\phi}\RP^2)\\
    \vect{v}^\text{arm} &= t \sqrt{1-\Lambda_\text{arm}} \boldsymbol{A}_\text{arm}^{-1} \vect{u}\\
    \boldsymbol{\mathcal{L}}_\text{arm} &= t \sqrt{\Lambda_\text{arm}} \boldsymbol{A}_\text{arm}^{-1}\\
    \boldsymbol{A}_\text{arm} &= \boldsymbol{1} -r \sqrt{1-\Lambda_\text{arm}} e^{2\ii\phi} \RP^2
\end{align}
\begin{align}
    \RP &= \begin{bmatrix}
        1 & 0\\
        -K & 1
    \end{bmatrix} \\
    \vect{u} &= 
    \frac{\sqrt{2K}}{h_\text{SQL}} \begin{bmatrix}
        0 \\
        1
        \end{bmatrix}\\
    K &=\frac{8 P_\text{cav} \omega_0}{mc^2\Omega^2}\\
    h_\text{SQL} &= \sqrt{\frac{8\hbar}{m \Omega^2 L^2}}
\end{align}

Introducing a homodyne readout $\vect{u}(\theta) = \begin{bmatrix}
    \cos(\theta) \\
    \sin(\theta)
\end{bmatrix} $, the single-sided quantum noise PSD in strain units is
\begin{align}
    \boldsymbol{S}_h^\text{QN} &= \frac{1}{|\vect{u}^T\vect{v}^h|^2}\vect{u}^T(\boldsymbol{T}^\text{in}\boldsymbol{S}^\text{in}\boldsymbol{T}^{\text{in}\dagger} + \boldsymbol{T}^\text{arm}\boldsymbol{T}^{\text{arm}\dagger} \nonumber\\
    & \qquad \qquad \qquad \qquad + \boldsymbol{T}^\text{SR}\boldsymbol{T}^{\text{SR}\dagger}  + \boldsymbol{T}^\text{vac}\boldsymbol{T}^{\text{vac}\dagger})\vect{u}\\
    &= S_h^\text{in} + S_h^\text{arm} + S_h^\text{SR} + S_h
    ^\text{vac}
\end{align}
\noindent where $S_h^\text{i}$ corresponds to the quantum noise coming from the i-th source in strain units.


%

\end{document}